\documentclass[aps,pra,twocolumn,amsmath,amssymb,showpacs]{revtex4-1}
\usepackage{graphicx}
\usepackage{amsmath}
\usepackage{bm}
\usepackage{xcolor}

\newcommand{\be}{\begin{equation}}  
\newcommand{\ee}{\end{equation}}
\newcommand{\ba}{\begin{array}}
\newcommand{\ea}{\end{array}}
\newcommand{\bea}{\begin{eqnarray}}
\newcommand{\eea}{\end{eqnarray}}

\newcommand{\bra}{\langle}
\newcommand{\ket}{\rangle}

\newcommand{\nn}{\nonumber}
\definecolor{wine}{rgb}{0.64, 0.21, 0.21}

\begin{document}

\title{Energetic advantages of non-adiabatic drives combined with non-thermal quantum states\\
%Non-adiabatic drives combined with non-thermal states can outperform adiabatic drives and shortcut-to-adiabaticity
} 

\author{Camille L. Latune}
\affiliation{Quantum Research Group, School of Chemistry and Physics, University of
KwaZulu-Natal, Durban, KwaZulu-Natal, 4001, South Africa, and\\
National Institute for Theoretical Physics (NITheP), KwaZulu-Natal, 4001, South Africa}

\date{\today}
\begin{abstract}
Unitary drivings of quantum systems are ubiquitous in experiments and applications of quantum mechanics and the underlying energetic aspects, particularly relevant in quantum thermodynamics, are receiving growing attention. We investigate energetic advantages in unitary driving obtained from initial non-thermal states. We introduce the non-cyclic ergotropy to quantify the energetic gains, from which coherent (coherence-based) and incoherent (population-based) contributions are identified. In particular, initial quantum coherences appear to be always beneficial whereas non-passive population distributions not systematically. Additionally, these energetic gains are accessible only through non-adiabatic dynamics, contrasting with the usual optimality of adiabatic dynamics for initial thermal states. Finally, following frameworks established in the context of shortcut-to-adiabaticity, the energetic cost related to the implementation of the optimal drives are analysed and, in most situations, are found to be smaller than the energetic cost associated with shortcut-to-adiabaticity. We treat explicitly the example of a two-level system and show that energetic advantages increase with larger initial coherences, illustrating the interplay between initial coherences and the ability of the dynamics to consume and use coherences.

\end{abstract}
%\pacs{}

\maketitle

{\it Introduction}.---Most quantum experiments and quantum technologies require manipulation of quantum systems' Hamiltonian. %like in quantum thermodynamics, quantum adiabatic computing, quantum annealing, and quantum metrology. 
Among the infinite variety of drivings realizing the desired Hamiltonian transformation, %starting and ending with the desired Hamiltonians, 
the least energy-consuming ones are of high interest for energy controled applications, like in thermodynamics but soon in quantum information processing and computation \cite{Strubell_2019,Asiani_2020}.
 %applications like quantum thermal engines \cite{}.
%it is often welcome to choose the least energy demanding, as for instance in the diverse thermodynamic applications in classical and quantum regime \cite{}. %, quantum engines \cite{}, ...
%The former option is particularly sought in thermodynamic applications like quantum engines \cite{Otto cycles} 
These least energy-consuming unitary evolutions are commonly associated with the well-known family of adiabatic drives. The traditional criterion for adiabaticity relies on the slow variation of the driving with respect to the velocity of the system's evolution \cite{Born_1928} (see also \cite{Teufel_2003,Allahverdyan_2005,Albash_2012} for recent reformulation and extension).
The energetic aspects and the origin of non-adiabaticity -the breakdown of adiabaticity-
 were recently shown to stem from the non-commutativity of the time dependent Hamiltonian \cite{Kosloff_2002,Feldmann_2003,Feldmann_2004}, giving rise to generation of quantum coherences and consequently extra energetic costs \cite{Plastina_2014} as well as {\it irreversible work} \cite{Deffner_2010,Francica_2019,Mohammady_2020}. 
%
%
%While non-adiabadicity and its associated quantum friction can be circumvented using techniques like shortcut-to-adiabaticity \cite{Demirplak_2003,Demirplak_2005,Demirplak_2008,Berry_2009}, these same techniques have an intrinsic energetic cost \cite{papers by Obinna}. 
Such manifestations of {\it quantum friction} \cite{Kosloff_2002,Feldmann_2003,Feldmann_2004} can be circumvented using techniques like shortcut-to-adiabaticity \cite{Demirplak_2003,Demirplak_2005,Demirplak_2008,Berry_2009}, widely applied in theoretic and experimental thermodynamics \cite{delCampo_2014,Deng_2013,Beau_2016,Abah_2019,Hartmann_2020,Dann_2020,Deng_2018}, adiabatic quantum computing \cite{Hegade_2021}, experimental state engineering \cite{Chen_2020}, and quantum information processing \cite{Santos_2020}. %CUT?? THIS LIST OF APPLICATIONS CAN BE DROPPED.

 Nevertheless, the above considerations and results are valid for initial thermal states. 
 Here, we focus on initial non-thermal states and the energetic consequences for driving operations.
 We show that non-adiabatic drives become energetically optimal, 
 highlighting the ongoing interplay between the initial coherences contained in the system and the capacity of the drive to consume coherences.
We introduce the concept of {\it non-cyclic ergotropy} to quantify the corresponding energetic gains. We also investigate the energetic cost required for the implementation of the optimal drives. % {\it Add some of these things in the abstract and cut it from here}.
Compared to shortcut-to-adiabaticity techniques, we show explicitly in an example with a two-level system that non-adiabatic drives combined with initial non-thermal states can bring higher energetic gains simultaneously with lower energetic costs.\\

%\section{Optimality of non-adiabatic drives and energetic gains}
  Let us consider the operation consisting in driving a quantum system $S$ from an initial Hamiltonian $H_i$ to a final one $H_f$, with their respective eigenvalues and eigenvectors denoted by $e^x_n$ and $|e^x_n\ket$, for $x=i,f$, in increasing order, $e^x_{n+1}\geq e^x_{n}$. We start the analysis by one of the central quantity of the problem: $E_f:={\rm Tr}(\rho_f H_f)$, the energy of the final state $\rho_f$, reached at the end of the driving. For a given arbitrary initial state $\rho_i$ of initial energy $E_i:={\rm Tr}(\rho_iH_i)$, there is an infinite variety of driving Hamiltonians $H(t)$ satisfying $H(t_i) = H_i$ and $H(t_f)= H_f$, leading to an infinity of different final energy. Independently of whether the driving operation injects energy in $S$ ($E_f\geq E_i$) or extracts energy from $S$ ($E_f\leq E_i$), the optimal drive, which is in fact not unique, has to minimise $E_f$, so that it minimises the energetic costs or maximises the energetic gains of the operation.
  Therefore, our first aim is to find the minimum $\tilde E_f:={\rm min}_{U\in {\cal U}} {\rm Tr}(U\rho_iU^{\dag}H_f)$, where ${\cal U}$ is the ensemble of unitary operations generated by drives $H(t)$ satisfying $H(t_i) = H_i$ and $H(t_f)= H_f$. As we will see in the following, ${\cal U}$ is indeed simply equal to the ensemble of all unitary transformations -- in other words, any unitary transformation can be expressed as a unitary transformation generated by a time dependent Hamiltonian satisfying $H(t_i) = H_i$ and $H(t_f)= H_f$.
 %which is indeed simply equal to the ensemble of all unitary transformations (it will appear clearly in the following). %; secondly, we are also interested in the explicit expression of at least one of the optimal drives.  

Since all unitarily accessible final states have necessarily the same entropy as $\rho_i$, one might first think of $\tilde E_f$ as the smallest energy over the ensemble of states of same entropy as $\rho_i$. Then, given that the state of smallest energy at fixed entropy is a thermal state, one would conclude that $\tilde E_f$ corresponds to the energy of $(\rho_{i})^{\rm th}_f$, the thermal state with respect to $H_f$ of same entropy as $\rho_i$.  
%Then, at first sight, the minimal final energy should be achieved when $\rho_f$ is equal to $\rho_{f/i}^{\rm th}$, the thermal state of same entropy as $\rho_i$.
 However, reminding that unitary evolutions conserve eigenvalues, $(\rho_{i})^{\rm th}_f$ cannot in general be reached unitarily, unless the eigenvalues $r_n$ of the initial state $\rho_i$ are equal to the populations of a thermal state of $H_f$, as highlighted in \cite{Allahverdyan_2004}. 
% can be achieved only if the eigenvalues $r_n$ of the initial state $\rho_i$, when ordered in decreasing order $r_{n+1}\leq r_n$, are equal to the populations of a thermal state of $H_f$, as highlighted in \cite{Allahverdyan_2004}. 
Therefore, the state of lower energy which is always achievable through unitary operations is not $(\rho_{i})^{\rm th}_f$ but $\widetilde{(\rho_i)}_f$, a state diagonal in the eigenbasis of $H_f$ with eigenvalues equal to $r_n$, 
\be\label{passive}
\widetilde{(\rho_i)}_f := \sum_n r_n |e^f_n\ket\bra e^f_n|,
\ee
where $r_{n+1} \leq r_n$. The associated minimal difference of energy is
\be\label{nce}
-{\cal E}_{\rm nc}:= {\rm Tr}\left[\widetilde{(\rho_i)}_f H_f\right] - {\rm Tr}( \rho_iH_i) = \sum_n r_n  e_n^f - {\rm Tr} (\rho_iH_i).
\ee

The state in \eqref{passive} belongs to the family of passive states \cite{Pusz_1978, Lenard_1978}, defined as follows. For a given Hamiltonian $H = \sum_n e_n |e_n\ket\bra e_n|$, where the energies are ordered in increasing order, $e_{n+1}\geq e_n$, a state $\rho$ is said to be {\it passive with respect to $H$} if: (i) it is diagonal in the energy eigenbasis $\{|e_n\ket\}_n$; (ii) it has decreasing populations, $p_{n+1}=\bra e_{n+1}|\rho|e_{n+1}\ket\leq p_n=\bra e_n|\rho|e_n\ket$. The violation of any of these two conditions leads to two different types of {\it non-passivity}: non-passivity stemming from populations when (ii) is not fulfilled, and non-passivity stemming from coherences when (i) is not fulfilled. These different physical origins of non-passivity will be used in the next paragraph. Of course, it is also possible to have non-passivity stemming from both populations and coherences when neither (i) nor (ii) is fulfilled. 
Finally, a famous example of passive states is the thermal states.   %In particular, thermal states are passive states, but, for systems of three or more energy levels, not all passive states are thermal states.}
 
In the context of cyclic work extraction, where the aim is to extract as much work as possible from a quantum state $\rho$ through time-dependent driving under the {\it cyclic} constraint $H(t_i)=H(t_f)=H$, it was shown in pioneering studies \cite{Allahverdyan_2004,Alicki_2013} that no work can be cyclically extracted from passive states with respect to $H$. 
For states which are not passive, the maximal amount of cyclically extractable work is called {\it ergotropy}. Contrarily to what one could have expected, the ergotropy is not directly related to the minimal energy difference $ -{\cal E}_{\rm nc}$ -- the relevant quantity in our problem.
We call the quantity ${\cal E}_{\rm nc}$ the {\it non-cyclic ergotropy} since it is related to non-cyclic operations $H_i\ne H_f$. In particular, contrasting with the ergotropy, the non-cyclic ergotropy can be positive or negative. When positive, it represents the maximal energy extractable from $\rho_i$ while realising the driving from $H_i$ to $H_f$. When negative, its absolute value represents the minimal energy needed to take the system from $H_i$ to $H_f$ when starting from $\rho_i$. Additionally, passive states with respect to $H_i$ are not always the states of smallest non-cyclic ergotropy, as shown in the following (neither are the passive states with respect to $H_f$). %SO, WHICH STATES ARE THE STATES OF SMALLEST non-cyclic ERGOTROPY??
Before continuing, a small note on the notations: $\tilde \sigma$ denotes the passive state of same entropy as $\sigma$ (also called the passive state of $\sigma$) with respect to $H_i$. $\widetilde{(\sigma)}_f$ denotes the passive state of $\sigma$ with respect to $H_f$.\\

{\it Necessity of incoherent and coherent non-adiabatic transformations}.---It should be emphasised that any dynamics leading to a final passive state is necessarily non-adiabatic if and only if the initial state is a non-passive state with respect to $H_i$, contrasting with the usual adiabatic dynamics required for initial thermal states \cite{Deffner_2010,Plastina_2014,Francica_2019,Mohammady_2020}. This can be easily seen by writing the initial state in its diagonal form. %(more details in Supplementary Material \cite{SM}). %$\rho_i=\sum_n r_n |r^i_n\ket \bra r_n|$. %where the unit vectors $|r^i_n\ket$ are the eigenvectors associated with the eigenvalues $r_n$ (decreasingly ordered).

 Furthermore, we notice that there are two kinds of non-adiabatic transformations: the incoherent ones, which generate transitions between different initial and final energy levels but do not generates coherences in the eigenbasis of $H_f$, and the coherent ones, which do generate coherences in the eigenbasis of $H_f$. 
 This finds an interesting parallel with the type of non-passive features -- with respect to $H_i$ -- initially present in $\rho_i$. %, which stem either from the populations or from coherences.
  If the initial state contains non-passive features stemming only from populations, non-adiabatic evolutions yielding $\widetilde{(\rho_i)}_f$ are incoherent (see Appendix \ref{appcoherent}). Alternatively, if the non-passivity of $\rho_i$ is coherence-based, evolutions yielding $\widetilde{(\rho_i)}_f$ are necessarily coherent non-adiabatic. This highlights the interplay between coherences contained in $\rho_i$ and the ability of the evolution to consume and use coherences.
  %(see more details in Appendix \ref{appcoherent}).
  
 This difference in the nature of the required transformation is mirrored in ${\cal E}_{\rm nc}$: the non-cyclic ergotropy can be decomposed in a sum of an incoherent, a passive and a coherent contributions, ${\cal E}_{\rm nc} = {\cal E}_{\rm nc}^{\rm inc} + {\cal E}_{\rm nc}^{\rm pas} + {\cal E}_{\rm nc}^{\rm coh} $. The incoherent contribution can be defined as ${\cal E}_{\rm nc}^{\rm inc} := {\rm Tr}(\rho_i H_i)  - {\rm Tr}\left[ \widetilde{\rho_{i|_D}}   H_i\right]$, where $\rho_{i|_D}:= \sum_n \bra e_n^i|\rho_i|e_n^i\ket |e_n^i\ket\bra e_n^i|$ is the ``diagonal cut" of $\rho_i$ and $\widetilde{\rho_{i|_D}}$ its associated passive state with respect to $H_i$. The passive contribution can be identified as ${\cal E}_{\rm nc}^{\rm pas} := {\rm Tr}\left[ \widetilde{ \rho_{i|_D}} H_i\right] - {\rm Tr}  \left[\widetilde{ (\rho_{i|_D})}_f H_f\right]$ where $ \widetilde{ (\rho_{i|_D})}_f$ is the passive state of $\rho_{i|_D}$ with respect to $H_f$. Finally, the coherent contribution can be identified as ${\cal E}_{\rm nc}^{\rm coh} = {\rm Tr} \left[\widetilde{ (\rho_{i|_D})}_f H_f\right] - {\rm Tr}\left[\widetilde{(\rho_i)}_f H_f\right]$. Additional technical details can be found in Appendix \ref{appcoherent}.
 %an incoherent contribution, a coherent contribution, and a passive contribution \cite{SM}. %(see Appendix \ref{appcoherent}). 
 This extends similar considerations presented in \cite{Francica_2019,Francica_2020} on ergotropy. \\ %However, for non-cyclic ergotropy, it is convenient to introduce an additional passive contribution, described in Appendix \ref{appcoherent}.\\

{\it Energetic gains}.---We are now in position of evaluating the energetic advantages in driving operations provided by non-passivity. These advantages are given by the amount of energy gained (or saved) thanks to the use of the best strategy starting from a non-thermal state compared to the best strategy starting from a thermal state of same energy. 
 %One can consider the amount of energy gained (or saved) using a non-thermal state combined with an optimal drive compared to starting from a thermal state and applying an adiabatic drive or a shortcut-to-adiabaticity. 
 As detailed in the following, such energy gain is directly given by the difference of non-cyclic ergotropies between the initial thermal and non-thermal states. 
%These gains are given by the difference of non-cyclic ergotropies between initial thermal states and initial non-thermal states of same energy. % between the non-cyclic ergotropy of initial thermal states and initial non-passive states. 

More precisely, for an initial thermal state, it is well-known, as mentioned in the introduction, that energetically optimal drives are either adiabatic (quasi-static), or use shortcut-to-adiabaticity techniques \cite{Demirplak_2003,Demirplak_2005,Demirplak_2008,Berry_2009}. Then, from the initial thermal state $\rho_i^{\rm th}= \sum_n p_{i,n}^{\rm th} |e_n^i\ket\bra e_n^i|$, where $p_{i,n}^{\rm th}:= Z^{-1}e^{-\beta e_n^i}$, $Z:={\rm Tr} (e^{-\beta H_i})$ and $\beta$ plays the role of the inverse temperature, such drives yield the final passive state $\widetilde{(\rho_i^{\rm th})}_f $, given by \eqref{passive} substituting $r_n$ by $p_{i,n}^{\rm th}$. Note that $\widetilde{(\rho_i^{\rm th})}_f $ is generally not a thermal state if the energy spectrum of $H_f$ is not ``proportional" to the one of $H_i$. The non-cyclic ergotropy, applicable also for initial thermal states, is reached by these optimal drives and is given by \eqref{nce} ${\cal E}_{\rm nc} = {\rm Tr}(\rho_i^{\rm th}H_i) - {\rm Tr} \left[\widetilde{(\rho_i^{\rm th})}_f H_f\right]= \sum_n p_{i,n}^{\rm th} (e_n^i-e^f_n) $.
%the non-cyclic ergotropy is obtained \textcolor{wine}{using \eqref{nce}, which gives ${\cal E}_{\rm nc} = {\rm Tr}(\rho_i^{\rm th}H_i) - {\rm Tr} \left[\widetilde{(\rho_i^{\rm th})}_f H_f\right]= \sum_n p_{i,n}^{\rm th} (e_n^i-e^f_n) $. In this situation where the initial state is thermal with respect to $H_i$, the final passive state $\widetilde{(\rho_i^{\rm th})}_f $, given by \eqref{passive} substituting $r_n$ by $p_{i,n}^{\rm th}$, is reached by usual adiabatic (quasi-static) drives or shortcut-to-adiabaticity \cite{Demirplak_2003,Demirplak_2005,Demirplak_2008,Berry_2009}. We emphasises that $\widetilde{(\rho_i^{\rm th})}_f $ is generally not a thermal state if $H_f$ is not proportional to $H_i$.}

Thus, the non-cyclic ergotropy difference, representing the energy difference between the best strategies starting either from a thermal state $\rho_i^{\rm th}$ or from a non-thermal $\rho_i$ of same energy, is given by
\bea\label{deltaE}
\Delta {\cal E}_{\rm nc} &:=& {\rm Tr} \left[\widetilde{(\rho_i^{\rm th})}_f H_f\right] - {\rm Tr}\left[\widetilde{(\rho_i)}_fH_f \right]\nn\\
&=& \sum_n (p^{\rm th}_{i,n}-r_n)e^f_n.
%&=& \sum_n [(p^{\rm th}_{i,n}-p^{\rm nth}_{i,n})e^f_n + (p^{\rm nth}_{i,n} -r_n)e^f_n]\nn\\
%&=&\sum_n (p^{\rm th}_{i,n}-p^{\rm nth}_{i,n})e^f_n + {\cal E}_f,
\eea
% for an initial non-thermal state $\rho_i$ of same energy as $\rho_i^{\rm th}$.

 Is $\Delta {\cal E}_{\rm nc}$ always positive? 
Quite surprisingly, the answer is no, contrasting with cyclic ergotropy. It means that, some thermal states have a larger non-cyclic ergotropy than some non-passive states of same energy, or in other words, more work can be extracted non-cyclicly from some thermal states than from some non-passive states of same energy. %This behaviour contrasts from what one is used to with cyclic ergotropy.  
%there are some thermal state from which we can extract more work (positive no-cyclic ergotropy) during the non-cyclic transformation $H_i \rightarrow H_f$, or, conversely, requiring less energy to realise the transformation $H_i \rightarrow H_f$ (negative non-cyclic ergotropy). 
We provide explicit examples in Appendix \ref{appexample}.

A general condition guaranteeing the positivity of $\Delta {\cal E}_{\rm nc}$ is given by the property of majorization. We recall that for any two density operators $\rho$ and $\sigma$, $\rho$ majorizes $\sigma$ %, denoted usually by $\rho \succ \sigma$, 
when \cite{Allahverdyan_2004,Gour_2015}
\be
\sum_{n=1}^k  r_n \geq \sum_{n=1}^k s_n
\ee
for all $k\geq 1$, where $r_n$ and $s_n$ are respectively the eigenvalues of $\rho$ and $\sigma$, in decreasing order. %{\it Recall some basic properties of majorization?}. 
Then, the positivity of $\Delta {\cal E}_{\rm nc}$ is guaranteed when $\rho_i$ {\it majorizes} $\rho_i^{\rm th}$, which can be seen using summation by part \cite{Allahverdyan_2004}, $\Delta {\cal E}_{\rm nc} = \sum_n  (p^{\rm th}_{i,n}-r_n)e^f_n
 = \sum_{k\geq1} (e^f_{k+1}-e^f_k)\sum_{n=1}^k (r_n - p^{\rm th}_{i,n}) \geq 0$.

%PUT IN THE SM??

In particular, this implies that coherence-based non-passivity {\it always} lead to positive $\Delta {\cal E}_{\rm nc} $, while this is not true for population-based non-passivity. This unexpected difference stems from the passive contribution to $\Delta {\cal E}_{\rm nc}$, which is zero for coherence-based non-passivity whereas it can take any sign -and in particular the negative one- for population-based non-passivity, Appendix \ref{appcons}.\\ %\ref{appcoherent}).\\
%in terms of the different energetic contributions to the non-cyclic ergotropy introduced above and explicitly detailed in Appendix \ref{appcoherent}. For coherence-based non-passivity, the passive contribution is the same for both the non-passive and thermal states. Since the coherent contribution is always positive, one concludes that coherences always bring energetic advantages. By contrast, the passive contribution from population-based non-passivity is not necessarily the same as the thermal state, and in particular can be smaller, leading to a smaller non-cyclic ergotropy.\\

We mention briefly an alternative figure of merit quantifying the energetic advantages stemming from the optimal driving itself. It simply consists in the energy gained or saved by applying an optimal drive to a given initial state $\rho_i$ instead of applying an adiabatic drive or a shortcut-to-adiabaticity. It is given by $G_{\rho_i}:= {\rm Tr} \left\{\left[U_{\rm ad}\rho_iU_{\rm ad}^{\dag} -\widetilde{(\rho_i)}_f\right]H_f \right\} =\sum_n (p^i_n - r_n)e_n^f$, where $U_{\rm ad}$ denotes the unitary transformation generated by the adiabatic drive or shortcut-to-adiabaticity and $p_n^i:=\bra e_n^i|\rho_i|e_n^i\ket$ are the populations in the initial energy eigenbasis. This quantity corresponds to the {\it cyclic} ergotropy of the state $U_{\rm ad}\rho_iU_{\rm ad}^{\dag}$ (and therefore also of $\rho_i$) with respect to $H_f$, and thus is always positive by contrast with $\Delta {\cal E}_{\rm nc}$. Note that for initial states with non-passivity stemming from coherences, as in the examples considered below, we have $\Delta {\cal E}_{\rm nc} = G_{\rho_i}$. \\  %%%Can we express $G_{\rho_i}$ in terms of the coherent, incoherent, and passive contributions?

{\it Upper bound and achievability}.---The non-cyclic ergotropy is naturally upper bounded by 
\bea
{\cal E}_{\rm nc} &=& {\rm Tr} (\rho_i H_i ) - {\rm Tr}\left[ \widetilde{(\rho_i)}_f H_f\right] \nn\\
&\leq& {\rm Tr} (\rho_i H_i) - {\rm Tr} \left[(\rho_i)_f^{\rm th} H_f\right]
\eea
where $ (\rho_i)_f^{\rm th} $ is the thermal state of $H_f$ of same entropy as $\rho_i$, already introduced previously. We denote by $\beta_i$ its inverse temperature. %such that $ \rho_f^{\rm th}(\beta_i) $ has the same entropy as $\rho_i$. %The above upper bound holds because thermal states are the states of minimal energy at a given entropy (and Hamiltonian). (Note that the upper bound can be expressed as ${\rm Tr} \rho_i H_i - {\rm Tr} \rho_f^{\rm th} H_f = \beta^{-1}S[\rho_i|\rho_i^{\rm th}(\beta)] + \beta^{-1}[F(\rho_i^{\rm th}(\beta)) - F(\rho_f^{\rm th}(\beta))]$, where $F(\rho)$ denotes the equilibrium free energy of the $\rho$.)
%The difference of non-cyclic ergotropy between an initial thermal state $\rho_i^{\rm th}$ and an initial non-thermal state $\rho_i^{\rm nth}$ of same energy 
The ergotropy difference \eqref{deltaE} is therefore upper bounded by
\bea\label{upperboundenc}
\Delta {\cal E}_{\rm nc} &\leq& {\rm Tr} \left[\widetilde{(\rho_i^{\rm th})}_f H_f\right]   - {\rm Tr}[ (\rho_i)_f^{\rm th} H_f]\nn\\ 
&&= \beta_i^{-1} \Delta S + \beta_i^{-1}S\left[\widetilde{(\rho_i^{\rm th})}_f |(\rho_i)_f^{\rm th}\right] 
\eea
where $\Delta S:= S(\rho_i^{\rm th}) - S(\rho_i)$ is the difference of Von Neumann entropy and is positive since $\rho_i^{\rm th}$ and $\rho_i$ have the same energy. 
This upper bound is automatically saturated when the final passive state $\widetilde{(\rho_i)}_f$ is a thermal state (and therefore equal to $ (\rho_i)_f^{\rm th}$). However, when $\widetilde{(\rho_i)}_f\ne(\rho_i)_f^{\rm th}$, the upper bound can still be saturated asymptotically by using many copies of the non-thermal state $\rho_i$, see Appendix \ref{appEq7}. 
This relies on the theorem shown in \cite{Alicki_2013}. Similarly, $G_{\rho_i}$ is upper bounded by $G_{\rho_i} \leq  {\rm Tr} \left\{\left[U_{\rm ad}\rho_iU_{\rm ad}^{\dag} -(\rho_i)_f^{\rm th}\right]H_f \right\} $, which can be saturated in the same conditions as Eq. \eqref{upperboundenc} thanks to \cite{Alicki_2013}. 
%which states that for any state $\rho$ and for $N$ going to infinity, there exist a unitary transformation $U_N$ (not unique) such that 
%\be
%\frac{1}{N}{\rm Tr} U_N \otimes^N \rho U_N^{\dag} H_N \underset{N\rightarrow \infty}{\rightarrow} {\rm Tr} \rho^{\rm th} H,
%\ee

As a result, any non-thermal features is energetically beneficial in the asymptotic limit of many copies, whereas for a single copy, non-thermality and even non-passivity are not sufficient to guarantee some energetic benefits with respect to initial thermal states -- only majorization is sufficient. \\% IMPROVE THIS SENTENCE!!!!

{\it Cost of driving}.---%In the previous paragraphs we showed that energetic advantages can be obtained in driving operations from non-passive and non-thermal states. %and the different contributions from populations and coherences. 
The remaining questions concern the existence, the explicit form, and the associated energetic cost of optimal drivings saturating the non-cyclic ergotropy. %We discuss this aspect in the following as well as the intrinsic energetic cost for their implementation.
For a given initial non-thermal state $\rho_i=\sum_n r_n|r^i_n\ket\bra r^i_n|$, we are looking for Hamiltonians $H(t)$ that generate the final unitary transformation $R=\sum_n e^{i\phi_n}|e_n^f\ket\bra r^i_n|$ with the constraints $H(t_i) = H_i$ and $H(t_f)=H_f$ at initial and final times $t_i$ and $t_f$. The phases $\phi_n$ can be chosen freely if one assumes an experimental setup able to control and adjust them, otherwise they will be left random. 
%We also have the freedom to chose the phase of each final eigenstate, represented by $\phi_n$.
% Note however that depending on the experiment, full or even partial control of such phase might not be possible and then becomes a random phase.

For arbitrary initial and final Hamiltonian $H_i$ and $H_f$ we define $H_0(t) := \lambda_i(t)H_i + \lambda_f(t)H_f$, where $\lambda_i(t)$ and $\lambda_f(t)$ are real positive functions such that $\lambda_i(t_i)=\lambda_f(t_f)=1$ and $\lambda_i(t_f)=\lambda_f(t_i)=0$. Besides these initial and final conditions, $\lambda_i(t)$ and $\lambda_f(t)$ can be chosen freely, in particular to suits experimental constraints. %(example of experimental realisations [PRA 99,062103] and [PRL,123,240601]). 

One can show (Appendix \ref{appdriving}) 
that a family of drivings reaching $\widetilde{( \rho_i)}_f$ is of the form $H(t)=H_0(t) + V(t)$ with $V(t)=-\hbar\dot f(t)U_0(t) \chi  U_0^{\dag}(t)$. We introduced $U_0(t):= e^{-\frac{i}{\hbar}{\cal T}\int_{t_i}^{t} duH_0(u)}$ as the unitary transformation generated by the original drive $H_0(t)$, ${\cal T}$ is the time-ordering operator, $\chi:=-i\ln \left[U_0^{\dag}(t_f)R\right]$ represents a kind of ``overlap" between the aimed transformation $R$ and the one actually generated by the original drive $H_0(t)$, and $f(t)$ is a real function which can be chosen freely besides the following conditions $f(t_i)=\dot f(t_i) =\dot f(t_f)=0$ and $f(t_f)=1$. This also shows that the ensemble ${\cal U}$ introduced in the beginning of the paper contains indeed all unitary evolutions since the above reasoning can be repeated for any unitary instead of $R$.

%THIS PARAGRAPH CAN BE REDUCED
The additional driving $V(t)$ seems energetically costless at first sight since it does not contribute explicitly to the total work, $\int_{t_i}^{t_f} {\rm d}u {\rm Tr} \left[\rho_u \frac{\rm d}{{\rm d}u} [H_0(u)+V(u)] \right]= {\rm Tr}(\rho_fH_f)-{\rm Tr}(\rho_iH_i)$. Still, there is a intrinsic energetic cost associated with the additional driving $V(t)$. This was pointed out in the context of shortcut-to-adiabaticity  \cite{Santos_2015, Santos_2016, Zheng_2016, Campbell_2017} and captured by the time-averaged norm of the additional Hamiltonian or instantaneous additional driving energy \cite{Abah_2017}. Note that the Hamiltonian norm is also shown to be the relevant quantity to express energetic cost in extended Landauer principle \cite{Deffner_2021}. % and can be related to the tradeoff revealed by quantum speed limit \cite{}, namely velocity of the evolution versus energy spent ({\it double check}).  \\
Following these energetic analysis, the energetic cost associated with the additional drive $V(t)$ is $w:=\frac{1}{\tau}\int_{t_i}^{t_f}dt ||V(t)||$, where $\tau:=t_f-t_i$ and  $||V(t)||$ is the Frobenius norm of $V(t)$, equal to 
\bea 
||V(t)|| := \left[{\rm Tr} \left[V(t)V^{\dag}(t)\right]\right]^{1/2}= \hbar|\dot f(t)|\left[{\rm Tr} \left(\chi \chi^{\dag}\right)\right]^{1/2}.\nn\\
\eea
%where $U_0:= e^{-i{\cal T}\int_{t_i}^{t} duH_0(u)} $. 

The relation defining $\chi$ can be re-written as $ e^{i\chi} = \sum_n e^{i\phi_n}|e_n^{i}\prime\ket\bra r^i_n|$ with $|e_n^{i}\prime\ket := U_0^{\dag}(t_f)|e_n^f\ket$. Since $\sum_n e^{i\phi_n}|e_n^{i}\prime\ket\bra r^i_n|$ is a unitary matrix, it is diagonalisable in the form $ \sum_n e^{i\theta_n} |u_n\ket\bra u_n|$, with $\theta_n \in [-\pi;\pi[$ and $|u_n\ket$ is the associated eigenvector. Then, a suitable choice is $\chi = \sum_n \theta_n |u_n\ket\bra u_n|$, implying $||V(t)|| =\hbar |\dot f(t)| \left(\sum_n \theta_n^2\right)^{1/2}$,
and an energetic cost equal to $w= \left(\sum_n \theta_n^2\right)^{1/2} \frac{\hbar}{\tau}\int_{t_i}^{t_f} dt |\dot f(t)|$. Since $\int_{t_i}^{t_f} dt |\dot f(t)| \geq 1$, with the inequality saturated when $\dot f(t) \geq 0$ for all $t \in [t_i;t_f]$, we have the following achievable lower bound 
\be 
w \geq w_{\rm min} := \frac{\hbar}{\tau}\left(\sum_n \theta_n^2\right)^{1/2}.
\ee

%THIS WHOLE PARAGRAPH CAN BE REDUCED. Additionally, the statement about control of the phases was already made above, so maybe it can be cut here or there.
The term $\left(\sum_n \theta_n^2\right)^{1/2}$ can depend on the choice of the phases $\phi_n$. If we assume that we have experimentally the full control of such phases, we can choose them in order to minimise $\left(\sum_n \theta_n^2\right)^{1/2}$. Otherwise, the phases are random and we will simply average $\left(\sum_n \theta_n^2\right)^{1/2}$ over all possible phases to obtain an average cost. Finally, note that $\left(\sum_n \theta_n^2\right)^{1/2}$ is upper bounded by $\pi \sqrt d$,
%The additional energetic cost $w$ is trivially upper bounded by $w\leq \pi d  \frac{1}{\tau}\int_{t_i}^{t_f} dt |\dot f(t)| $, 
where $d$ is the dimension of the system.   \\

\begin{figure}
\centering
\includegraphics[width=7cm, height=4.5cm]{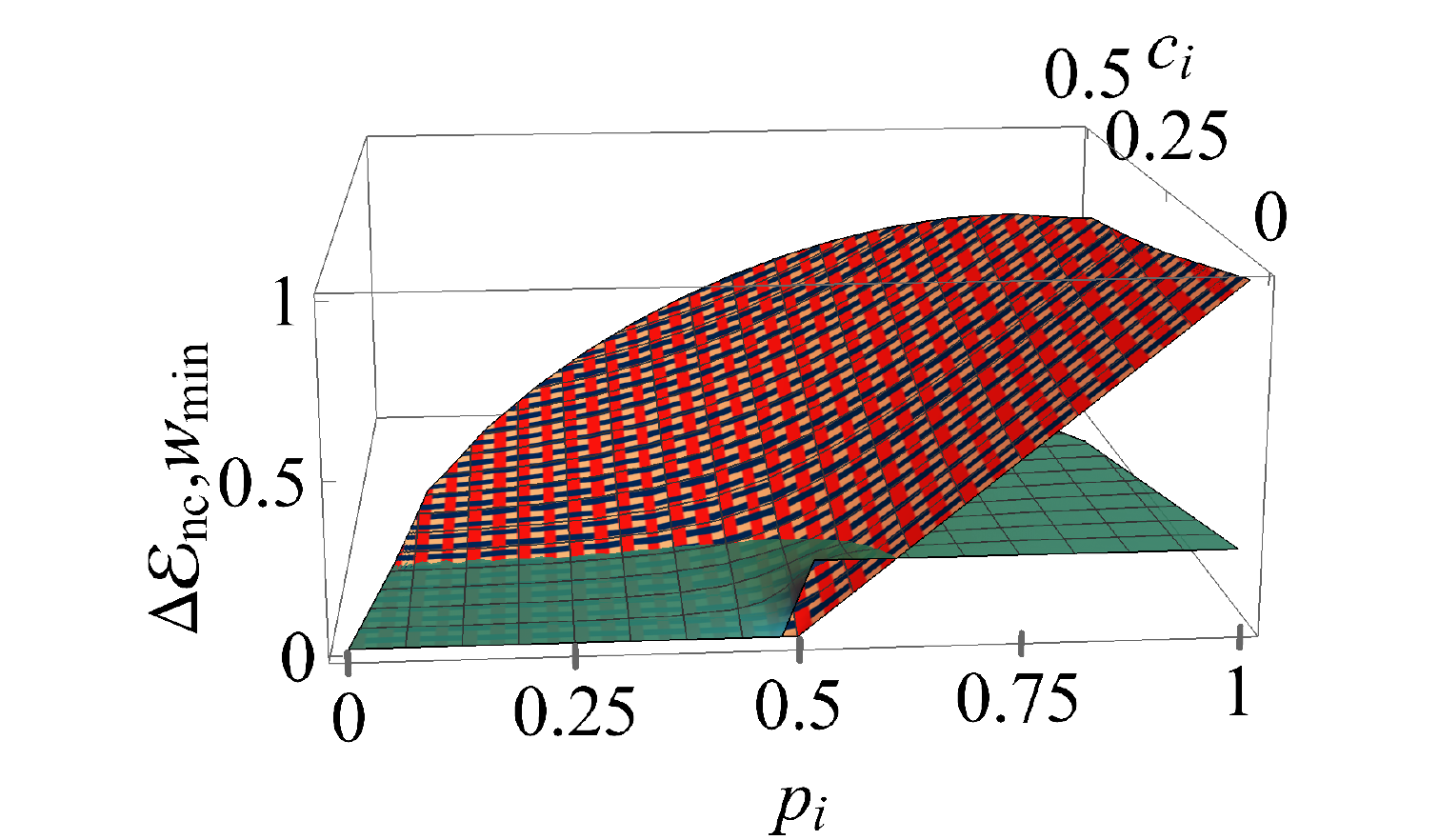}
\includegraphics[width=9.2cm, height=4.8cm]{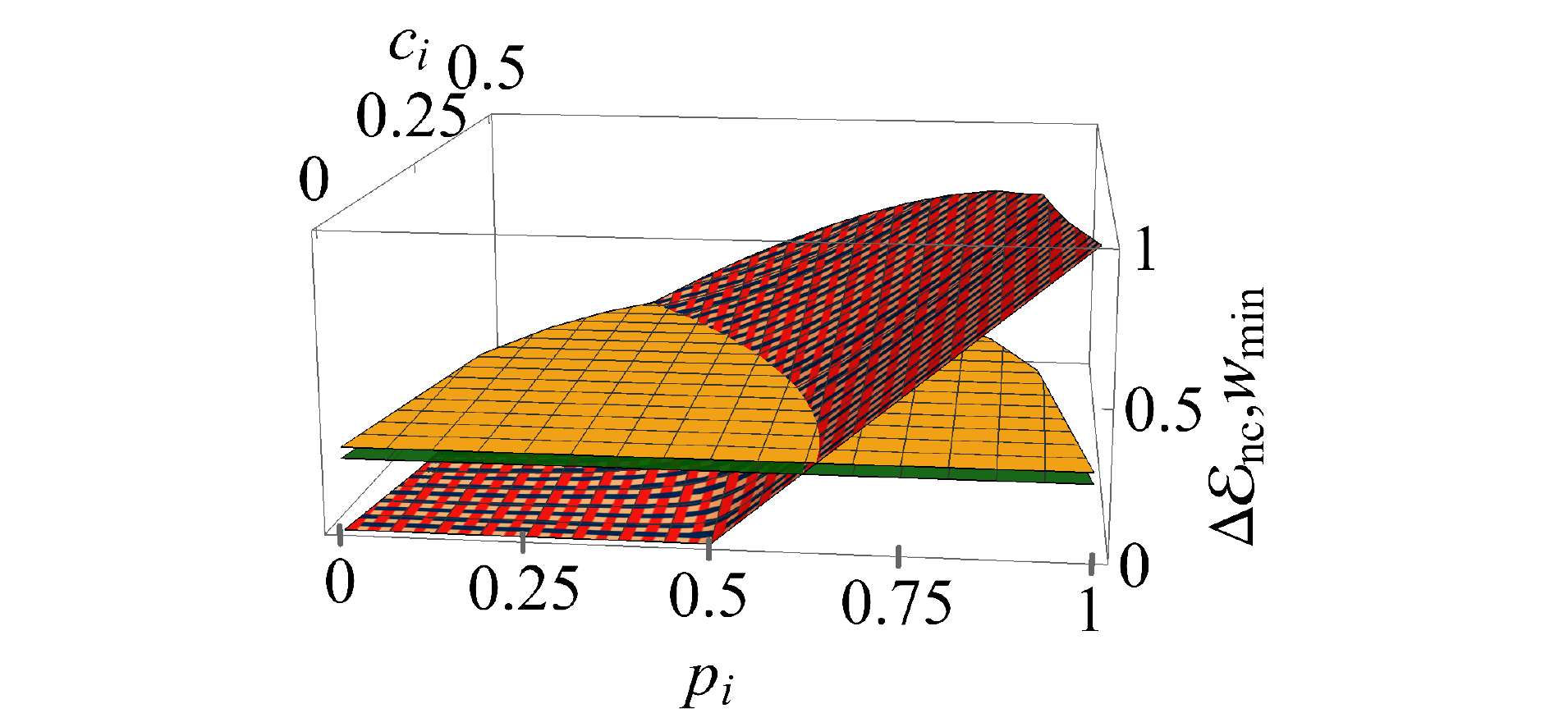}
\caption{Top Panel: Plot of the energy gain $\Delta {\cal E}_{\rm nc} =G_{\rho_i}$ (texturised surface) and the energetic cost $w_{\rm min}$ (solid green surface) both in unit of $\hbar\lambda_f\omega$, in function of the initial population $p_i\in[0;1]$ and coherence $|c_i|\in[0;\sqrt{p_i(1-p_i)}]$. We assume a driving velocity slower than the free evolution, setting $ 1/\tau =\lambda_f\omega/10$, which allows one to convert $\hbar/\tau$ in unit of $\hbar\lambda_f\omega$. Bottom Panel: Same plot with the upper bound $0.89\frac{\hbar\pi}{\tau}$ (yellow horizontal plane) and the lower bound $0.77\frac{\hbar\pi}{\tau}$ (green horizontal plane just below the yellow plane) of the average energetic cost $\overline{w_{\rm min}}$ (when no experimental control of the phases is available). }
\label{TLS}
\end{figure}

{\it Illustrative examples}.--- So far, we showed that non-thermal features can be used to gain or save energy in driving operations. %In particular, coherences appeared to always bring energetic advantages when combined with optimal drives, while non-thermal features in general might require several copies of the state to activate their energetic benefits. 
On the other hand, we also saw that there is an intrinsic energetic cost associated with optimal drives. Then, comes the following question: how large the energetic gains and the intrinsic energetic costs can be? We answer this question in two practical examples involving a two-level system. In the first example, we analyse the situation where $H_i$ is proportional to $H_f$ and compare the energetic gains provided by quantum coherences versus the energetic costs associated with the optimal drives.
In the second example, we consider the more general situation where $H_i$ and $H_f$ are not proportional. Then, the bare dynamics $U_0(t)$ is not adiabatic and shortcuts-to-adiabaticity are needed in order to reach $\widetilde{(\rho_i^{\rm th})}_f$ when starting from a thermal state $\rho_i^{\rm th}$. Thus, in this second example, beyond evaluating the energetic gains provided by quantum coherences, we also compare the intrinsic energetic costs between the optimal drives and shortcut-to-adiabaticity. 
%As mentioned above, both figures of merit of the energetic gains $\Delta {\cal E}_{\rm nc}$ and $G_{\rho_i}$ are equal in these examples.

{\it Example 1}.---We start by analysing the simple but common situation of a two-level system driven by a drive of the form $H_0(t):= \lambda(t)\hbar \frac{\omega}{2} \sigma_z$, which is naturally adiabatic since $[H_0(t),H_0(t')]=0$ for all $t$ and $t'$ \cite{Kosloff_2002,Feldmann_2003}. 
%We start by analysing a simple but common situation of a two-level system with $H_i$ and $H_f$ being proportional and the original driving $H_0(t):= \lambda(t)\hbar \frac{\omega}{2} \sigma_z$ is naturally adiabatic. 
The time-dependent parameter takes the initial and final positive value $\lambda(t_i):=\lambda_i$ and $\lambda(t_f):=\lambda_f$, and $\sigma_z$ denotes the z-Pauli matrix, with $|1\ket$ and $|0\ket$ the excited and ground states, respectively.
% $H_i= \lambda_i\hbar \frac{\omega}{2} \sigma_z$ being proportional to $H_f= \lambda_f\hbar \frac{\omega}{2} \sigma_z$, where $\lambda_i$ and $\lambda_f$ are positive numbers, $\sigma_z$ denotes the z-Pauli matrix, and $|1\ket$ and $|0\ket$ denote the excited and ground states, respectively. 
 A general initial non-thermal state is of the form $\rho_i = \begin{pmatrix}
    p_i & c_i \\
    c_i^{*} & 1-p_i \\
\end{pmatrix}$ in the basis $\{|1\ket,|0\ket\}$.  The associated non-cyclic ergotropy is given by \eqref{nce} with the eigenvalues $r_1$ and $r_0$ functions of $c_i$ and $p_i$ (expressions detailed in Appendix \ref{appexTLS}). The minimum energetic cost associated with the family of optimal drivings $V(t)$ is, according to the previous paragraph, given by %(see Appendix \ref{appexTLS})
 $w_{\rm min} = \frac{\sqrt{2}\hbar}{\tau}{\rm arctan} \frac{p_i-r_1}{r_0-p_i}$. %={\rm arctan}\left(\frac{\sqrt{1+\frac{|c|^2}{(1/2-p)^2}}-1}{\sqrt{1+\frac{|c|^2}{(1/2-p)^2}}+1}\right)^{1/2}$.
 
On the other hand, the initial thermal state of same energy as $\rho_i$ is simply $\rho_i^{\rm th} = {\rm diag}(p_i,1-p_i)$, and the original drive $H_0(t)$ is already adiabatic as commented above.
%final passive state $\widetilde{(\rho_i^{\rm th})}_f$ is naturally reached by the bare dynamics $U_0(t)$ generated by $H_0(t)$. 
Then, the energetic gain provided by the coherences $c_i$ is given by the non-cyclic ergotropy difference \eqref{deltaE}, which gives here 
%
%
%The associated eigenvalues and eigenstates are functions of $c_i$ and $p_i$, and substituting their expression %(Appendix \ref{appexTLS}) 
%in \eqref{deltaE}, the achievable energetic gain provided by $\rho_i$ \textcolor{wine}{compared to the thermal state of same energy (here, simply $\rho_i^{\rm th} = {\rm diag}(p_i,1-p_i)$)}, is \cite{SM}
% We re-write it in the form $\rho_i= r_1|r_1\ket\bra r_1| + r_0|r_0\ket\bra r_0|$ in term of the eigenvalues and eigenstates, whose expression depends on $p_i$ and $c_i$ and can be found in Appendix \ref{appexTLS}. The energetic gain \eqref{deltaE} in the driving operation $H_i \rightarrow H_f$ brought by the non-thermal state $\rho_i$ is up to
\bea
\Delta {\cal E}_{\rm nc} %&=& \lambda_f \frac{\omega}{2}(p_i-r_1) -\lambda_f \frac{\omega}{2}(1-p_i -r_0)\nn\\
&=& \hbar\lambda_f \omega \left[\sqrt{\left(1/2-p_i\right)^2 + |c_i|^2} -1/2+p_i\right]\geq 0,\nn
\eea
and is also equal to the alternative figure of merit $G_{\rho_i}$. 
%where the second line is 
%obtained using the expression of $r_1$ and $r_0$ given in the Appendix \ref{appexTLS}. %Note that $\Delta E$ does not depend on the phase of the coherences, as it should be. %EXPLAIN MORE THAT???
%Note that for two-level systems we always have $\Delta_{\rm dr} E = \Delta E$ since states of same energy have necessarily equal populations. 

%The minimum energetic cost associated with the family of optimal drivings $V(t)$ is given by \cite{SM} %(see Appendix \ref{appexTLS})
% $w_{\rm min} = \frac{\sqrt{2}\hbar}{\tau}{\rm arctan} \frac{p_i-r_1}{r_0-p_i}$. %={\rm arctan}\left(\frac{\sqrt{1+\frac{|c|^2}{(1/2-p)^2}}-1}{\sqrt{1+\frac{|c|^2}{(1/2-p)^2}}+1}\right)^{1/2}$.
Fig. \ref{TLS} (a) presents a plot of $\Delta {\cal E}_{\rm nc} = G_{\rho_i}$ and $w_{\rm min}$ assuming a driving velocity slower than the free evolution, $ 1/\tau =\lambda_f\omega/10$. One can see that the energetic gain is always larger than the cost for large initial coherences, as long as $p_i\geq 0.025$ (obtained numerically, not visible on the figure). Note that, rigorously speaking, $\Delta {\cal E}_{\rm nc}$ becomes ill-defined for $p_i\geq 1/2$ because then $\rho_i^{\rm th}$ is not anymore passive (negative temperature). Still, we can use $G_{\rho_i}$ to consider the energetic gain beyond $p_i =1/2$. Then, the sudden step in the driving cost at $p_i = 1/2$ happens because at this point the populations become inverted. This implies that, for $c_i=0$, optimal drives must swap the two eigenstates $|0\ket$ and $|1\ket$, whose cost is precisely $\frac{\pi\hbar}{\tau\sqrt{2}}$.
 %However, in situation where $\lambda_f \omega \gg 1/\tau$, the initial non-thermal state starts bringing advantage in most situations, even taking into account the driving cost.
  If the experimental setup does not offer control of the phases $\phi_1$ and $\phi_2$, the average energetic cost $\overline{w_{\rm min}}$ takes value between $0.77\pi\frac{\hbar}{\tau}$ and $0.89\pi\frac{\hbar}{\tau}$, displayed in Fig. \ref{TLS} (b).

\begin{figure}
\centering
~~(a)\includegraphics[width=8cm, height=4.6cm]{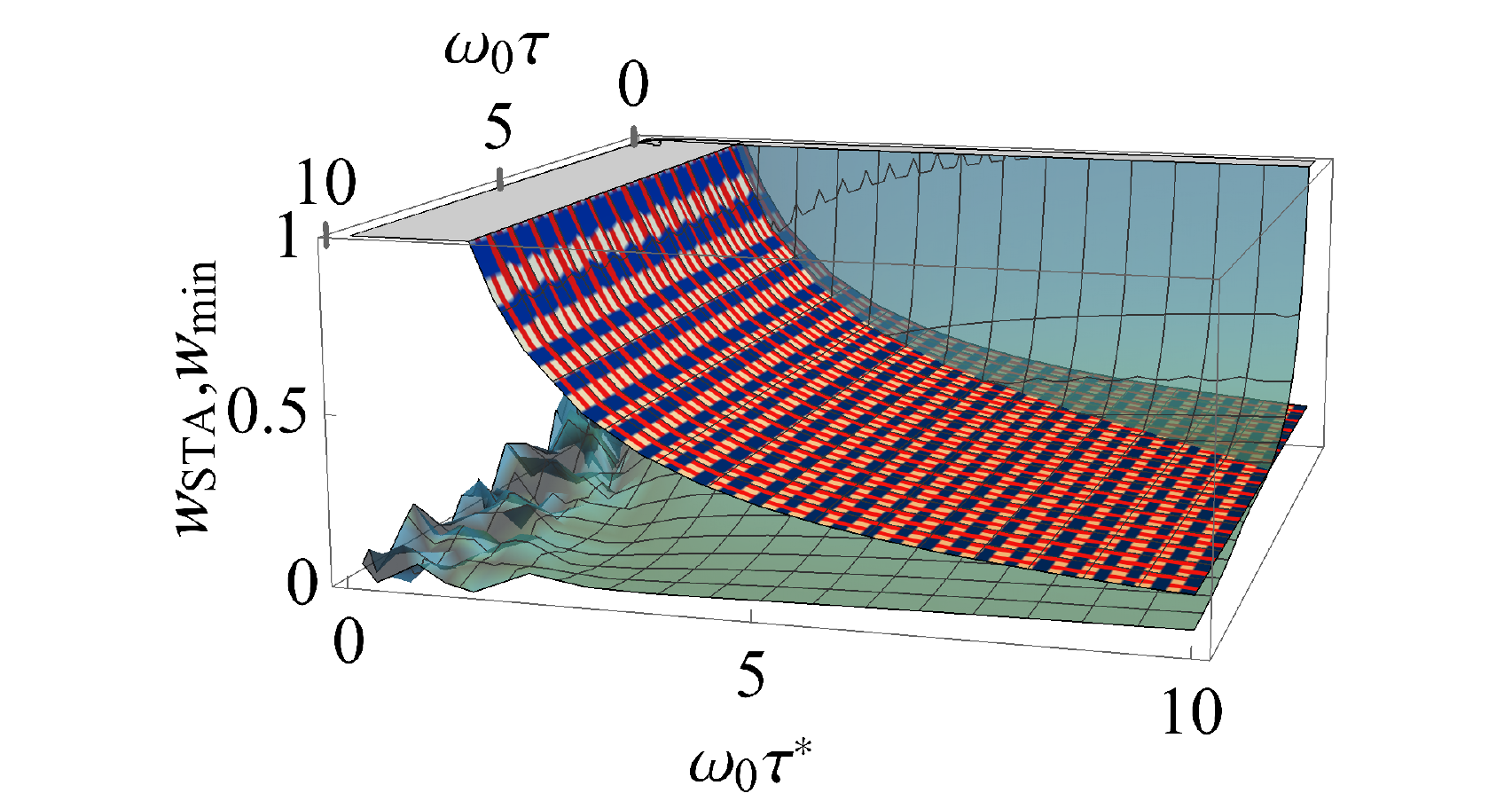}\\
~~~~(b)\!\!\!\!\!\!\!\!\!\!\!\!\!\!\!\!\!\includegraphics[width=9.5cm, height=4.3cm]{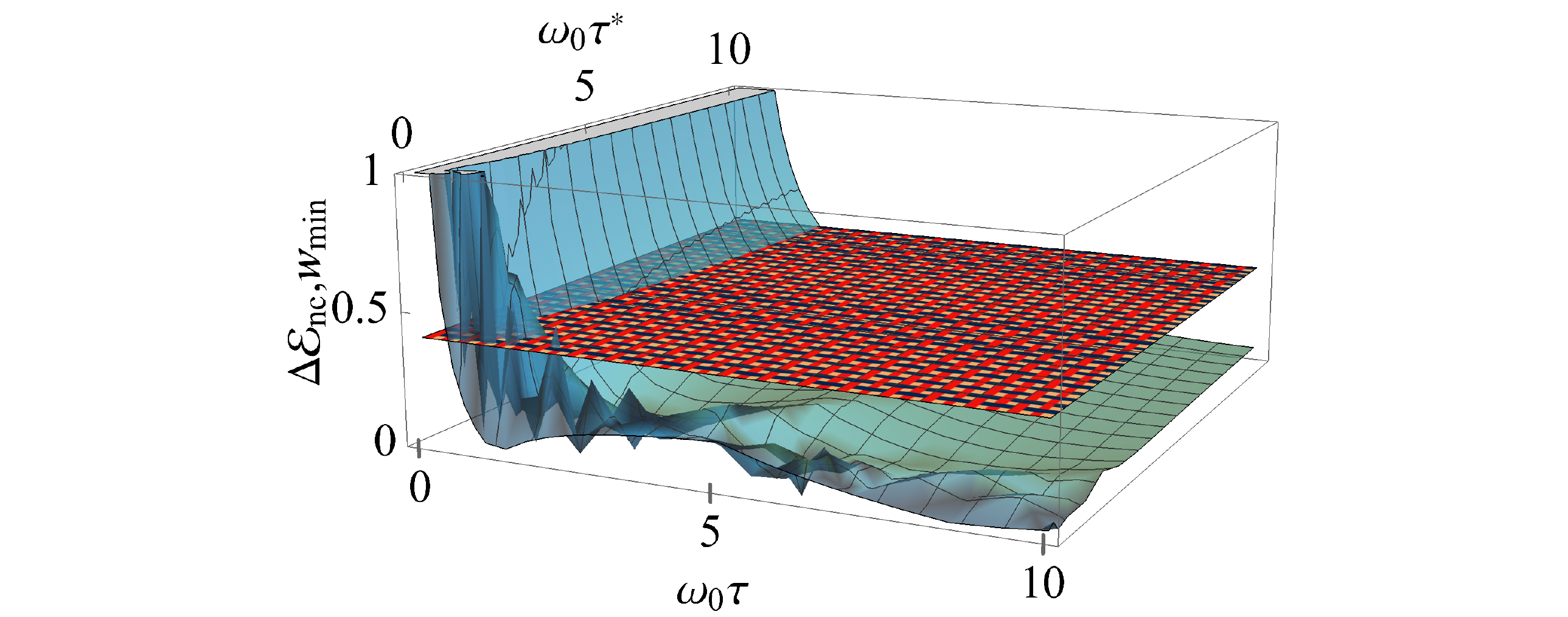}\\
(c)\includegraphics[width=6.5cm, height=4.5cm]{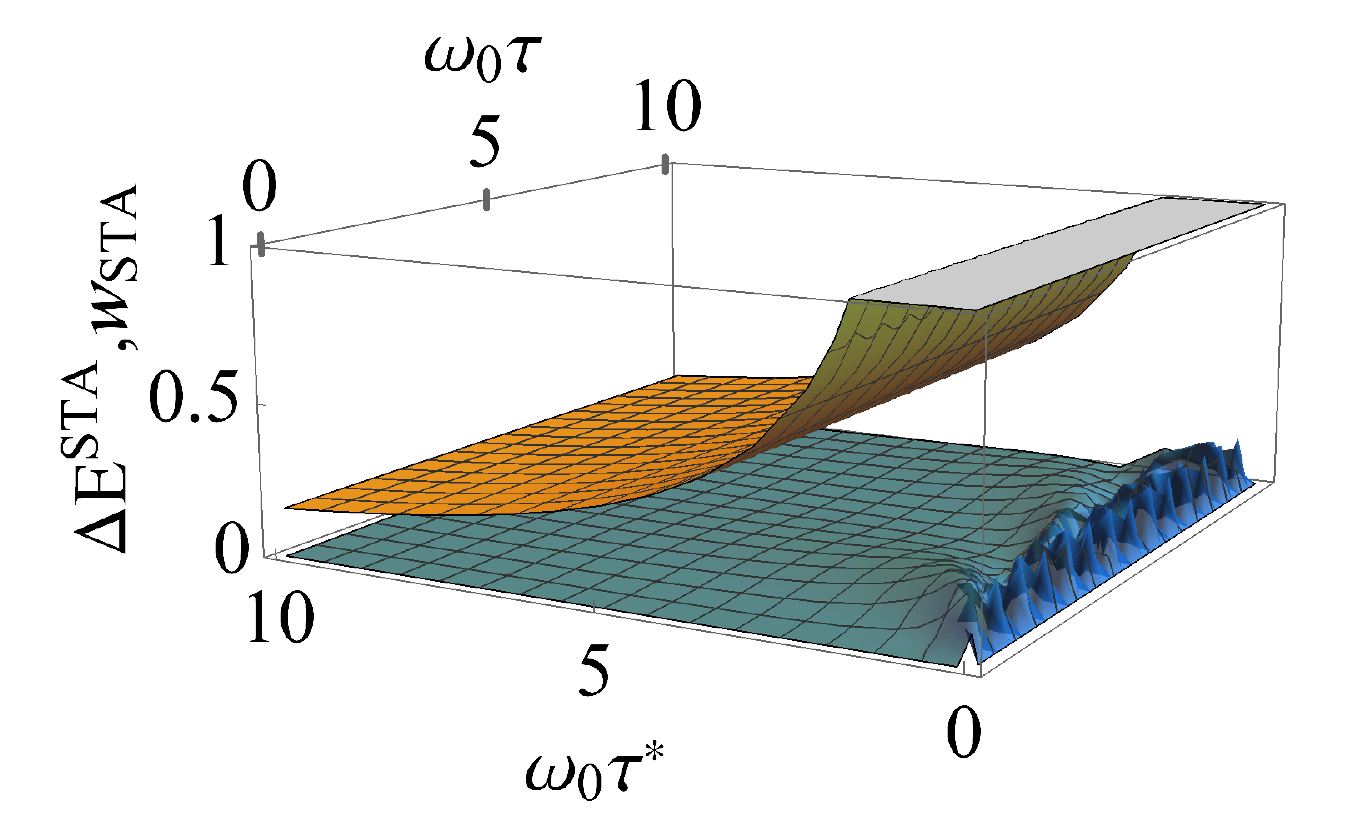}
\caption{Plots of  (a) $w_{\rm STA}$ (texturised surface) and $w_{\rm min}$ (blue solid surface); (b) $\Delta {\cal E}_{\rm nc} = G_{\rho_i}$ (horizontal texturised plane) and $w_{\rm min}$ (blue solid surface); and (c) $\Delta E^{\rm STA}$ (lower blue surface) and $w_{\rm STA}$ (upper yellow surface). All plots are in unit of $\hbar\omega_0$ and in function of $\omega_0\tau$ and $\omega_0\tau^{*}$, for $p_i=0.4$ and $c_i = \sqrt{0.4\times0.6}$. }
\label{comparisonmain}
\end{figure}

 {\it Example 2}.---We consider now the same system but with $H_i:=\frac{1}{2}\hbar\omega_0 \sigma_z$ and $H_f:= \frac{\hbar\omega_f}{2}\sigma_z + \frac{\hbar\epsilon_f}{2}\sigma_x$ not commuting, implying that $H_0(t)$ is necessarily non-adiabatic. We focus on the following family of driving,
 $H_0(t) = \frac{1}{2}\hbar\omega(t) \sigma_z + \frac{1}{2}\hbar\epsilon(t)\sigma_x$, with $\omega(t_i)=\omega_0$, $\epsilon(t_i)=0$, $\omega(t_f)=\omega_f$, and $\epsilon(t_f)=\epsilon_f$. In order to allow for analytic treatment of the dynamics, we assume that the time-dependent frequencies are such that $\mu := \frac{\dot \omega(t) \epsilon(t) - \dot \epsilon(t) \omega(t)}{\Omega^3(t)}$ is constant \cite{Feldmann_2003, Dann_2020}, where $\Omega(t) := \sqrt{\omega^2(t)+\epsilon^2(t)}$. %Since $[H_0(t),H_0(t')]\ne0$, the generated dynamics is non-adiabatic \cite{Kosloff_2002, add papers}. 
 This includes for instance time dependent frequencies of the form $\omega(t)=\omega_0\cos[\pi (t-t_i)/2\tau^{*}]$ and $\epsilon(t)=\omega_0\sin[\pi (t-t_i)/2\tau^{*}]$, commonly used experimentally \cite{Peterson_2019}. Such choice implies $\mu = - \pi/(2\omega_0\tau^{*})$, $\omega_f = \omega_0\cos[\pi \tau/2\tau^{*}]$ and $\epsilon_f=\omega_0\sin[\pi \tau/2\tau^{*}]$, whose exact value will depend on one's choice of $\tau^{*}$. In particular, for $\omega_0\tau^{*} \rightarrow \infty$, the adiabatic parameter \cite{Dann_2020} $\mu$ goes to zero, indicating adiabaticity, while $|\mu|\rightarrow \infty$ for $\omega_0\tau^{*} \rightarrow 0$, indicating strong non-adiabaticity.

 Since $H_0(t)$ is non-adiabatic (at least for finite $\omega_0\tau$), optimal drives for initial thermal states $\rho_i^{\rm th}$ use shortcut-to-adiabaticity \cite{delCampo_2014,Deng_2013,Beau_2016,Abah_2019,Hartmann_2020,Dann_2020,Deng_2018}. It consists in adding an extra drive, like for instance the so-called counter-diabatic drive $V_{\rm CD}(t)$ \cite{Demirplak_2003,Demirplak_2005,Demirplak_2008,Berry_2009}, whose aim is to suppress generation of coherences and level transitions. Then, for an initial thermal state $\rho_i^{\rm th} = p_i|1\ket\bra1| + (1-p_i)|0\ket\bra 0|$, the addition of the counter-diabatic drive $V_{\rm CD}(t)$ yields the final passive state $\widetilde{(\rho_i^{\rm th})}_f := p_i |e_1^f\ket\bra e_1^f| + (1-p_i)|e_0^f\ket\bra e_0^f|$ (which happens to be also a thermal state since there is only two energy levels), and reaches the non-cyclic ergotropy equal to ${\cal E}_{\rm nc}^{\rm STA} = E_i - {\rm Tr} \left[\widetilde{(\rho_i^{\rm th})}_f H_f\right]$.
  The expression of $V_{\rm CD}(t)$ tailored for our problem is provided in Appendix \ref{appcostSTA}
  (see also \cite{Feldmann_2003,Dann_2020}) as well as the derivation of the associated energetic cost $w_{\rm STA}$ according to the criteria discussed above. We find  $w_{\rm STA}  = |\mu|\frac{\hbar\bar{\Omega}}{\tau}$, with $\bar\Omega := \int_{t_i}^{t_f} dt \Omega(t)$.
 % = \hbar[\omega_0-\Omega_f](p_i-1/2)$. 

It is also interesting to estimate how much energy is indeed gained or saved thanks to the shortcut-to-adiabaticity, and compare it to $w_{\rm STA}$. This requires to compute the energy of the final state $\rho_f$ we would obtain using only the bare drive $H_0(t)$. We obtain ${\rm Tr} (\rho_fH_f)  =\Omega_f (p_f-1/2)$, where $\Omega_f:=\Omega(t_f)$ and $p_f=\bra e_1^f|\rho_f|e_1^f\ket$ is the population in the final excited state (analytical expression provided in Appendix \ref{appnonadiab}).
Thus, the energetic gain associated with the counter-adiabatic drive is 
%\textcolor{wine}{Additionally, the energy involved in the bare drive $H_0(t)$ is different from the non-cyclic ergotropy and is equal to $\Delta E_{\rm bare}:=E_i - {\rm Tr} (\rho_fH_f) $. %= \hbar\omega_0(p_i-1/2) -\Omega_f(p_f-1/2)$,
% Thus, the counter-diabatic drive provides the following energetic gain}
  %the difference of the non-cyclic ergotropies, $\Delta {\cal E}_{\rm nc}^{\rm STA} := {\cal E}_{\rm nc}^{\rm STA} -{\cal E}_{\rm nc} =  {\rm Tr} (\rho_f - \rho_f^{\rm th})H_f =\Omega_f (\frac{1}{2}-p_i)\frac{\mu^2(1-\nu_c)}{1+\mu^2}$. 
%$\Delta E^{\rm STA}:=  {\cal E}_{\rm nc}^{\rm STA} -\Delta E_{\rm bare} = {\rm Tr} (\rho_fH_f) -{\rm Tr} (\rho_f^{\rm th}H_f) =\hbar\Omega_f (\frac{1}{2}-p_i)\frac{\mu^2(1-\nu_c)}{1+\mu^2}$, 
$\Delta E^{\rm STA}:=  {\rm Tr} (\rho_fH_f) -{\rm Tr}  \left[\widetilde{(\rho_i^{\rm th})}_fH_f\right] =\hbar\Omega_f (\frac{1}{2}-p_i)\frac{\mu^2(1-\nu_c)}{1+\mu^2}$, with $\nu_c := \cos\bar\Omega\sqrt{1+\mu^2}$. %and $\bar\Omega := \int_{t_i}^{t_f} dt \Omega(t)$. 
Note that the energetic gain $\Delta E^{\rm STA}$ is positive only for initial positive temperature ($p_i\leq1/2$). This is because thermal states of negative temperature are non-passive, and then shortcut-to-adiabaticity techniques stop being optimal.

 We now focus on the non-cyclic ergotropy achieved by an arbitrary non-passive state $\rho_i = \begin{pmatrix}
    p_i & c_i \\
    c_i^{*}& 1-p_i \\
\end{pmatrix}$ (in the initial eigenbasis). Using the family of optimal drives $V(t)$ we can achieve the optimal final state $\widetilde{(\rho_i)}_f  =r_1 |e_1^f\ket\bra e_1^f| + r_0|e_0^f\ket\bra e_0^f|$, %CAN BE REDUCED!!!!
%= \begin{pmatrix}
 %   r_1 & 0 \\
 %   0& r_0 \\
%\end{pmatrix}$
 with the expressions of $r_1$ and $r_0$ as in the example 1. 
 The energetic gain with respect to the performance of shortcut-to-adiabaticity technique applied to initial thermal states is $\Delta {\cal E}_{\rm nc} %= {\cal E}_{\rm nc}^{\rm np}- {\cal E}_{\rm nc}^{\rm STA}
  = {\cal E}_{\rm nc} - {\cal E}_{\rm nc}^{\rm STA} = \hbar\Omega_f \left[\sqrt{(\frac{1}{2}-p_i)^2+|c_i|^2} - (\frac{1}{2}-p_i)\right]\geq 0$, same as in example 1, which is also equal to the alternative figure of merit $G_{\rho_i}$. %, and the energetic gain with respect to the bare driving $H_0(t)$ .
   However, differently from the example 1, one can now compare the energetic costs $w_{\rm STA}$ and $w_{\rm min}$, which provides a fairer comparison of performances between initial thermal states and non-passive states. % THE SECOND LINE CAN BE CUT % take into account $w_{\rm STA}$, the extra energetic cost required to reach the thermal state $\rho_f^{\rm th}$, and compare it to the energetic cost $w_{\rm min}$ associated with $V(t)$.
   The values taken by $w_{\rm min}$ now depends on some phases like the phase of the initial coherence $c_i$ (see technical details in Appendix \ref{appfamily}).
    We find that $w_{\rm min}$  belongs to the interval $\frac{\sqrt{2}\hbar}{\tau}|\theta_{1,{\rm min}} - \theta_{2,{\rm min}}| \leq w_{\rm min} \leq \frac{\sqrt{2}\hbar}{\tau}\pi$, where $\theta_{1,{\rm min}} := {\rm arctan} \sqrt{\frac{p_i-r_1}{r_0-p_i}} $ corresponds to the minimal energetic cost in example 1 and  $\theta_{2,{\rm min}} := {\rm arctan}\frac{|\mu|\sqrt{1-\nu_c}}{\sqrt{2+\mu^2(1+\nu_c)}}$ is only due to the non-adiabaticity of $U_0(\tau)$. Importantly, if one has experimental control of the phases, the minimal energetic cost $\frac{\sqrt{2}\hbar}{\tau}|\theta_{1,{\rm min}} - \theta_{2,{\rm min}}|$ can always be achieved.

%Their difference is $\Delta E_{\rm STA} - \Delta E = \Omega_f \left[\sqrt{(\frac{1}{2}-p_i)^2+|c_i|^2} - (\frac{1}{2}-p_i)\right]\geq 0$.
In Fig. \ref{comparisonmain} (a), we compare these energetic costs providing a plot of $w_{\rm STA}$ for $p_i=0.4$ and $w_{\rm min}$ for $p_i=0.4$ and $c_i=\sqrt{0.4\times0.6)}$. We use the specific form of the time dependent frequencies mentioned above, implying $\Omega(t)=\omega_0$, $\overline{\Omega} = \omega_0 \tau$, and $w_{\rm STA} = \frac{\hbar\pi}{2\tau^{*}}$. % $\omega(t)=\omega_0\cos[\pi (t-t_i)/2\tau^{*}]$ and $\epsilon(t)=\omega_0\sin[\pi (t-t_i)/2\tau^{*}]$, 
Thus, the plots are in unit of $\hbar\omega_0$ and in function of $\omega_0\tau^{*}$, directly related to the level of non-adiabaticity, and $\omega_0\tau$, related to how fast is the driving with respect to the free evolution of the system. One can see that $w_{\rm min}$ is smaller than $w_{\rm STA}$ for most values of $\omega_0\tau$ and $\omega_0\tau^{*}$. In particular, while $w_{\rm STA}$ diverges for high non-adiabaticity, $w_{\rm min}$ remains finite. However, $w_{\rm min}$ diverges for very fast drives. Interestingly, one can show that this divergence of $w_{\rm min}$ is only due to the contribution from $\theta_{1,{\rm min}}$. In particular, for $c_i=0$, $w_{\rm min}$ remains finite and strictly smaller that $w_{\rm STA}$, meaning that the driving $V(t)$ is more performant than shortcut-to-adiabaticity.

  In Fig. \ref{comparisonmain} (b), we compare the energetic gain $\Delta {\cal E}_{\rm nc}={\cal E}_{\rm nc} - {\cal E}_{\rm nc}^{\rm STA} $ (also equal to $G_{\rho_i}$) brought by the optimal drives with its energetic cost $w_{\rm min}$. 
 In Fig. \ref{comparisonmain} (c), we compare the energetic gain $\Delta E^{\rm STA}$ brought by shortcut-to-adiabaticity with its energetic cost $w_{\rm STA}$. One can see that the performances of the optimal drives are significantly better than the performances of shortcut-to-adiabaticity.
 Additional plots in function of the more general parameters $\overline{\Omega}$ and $|\mu|$ are available in Appendix \ref{appaddplots}, allowing us to conclude that the above tendency remain valid in more general settings. \\

{\it Conclusion}.---
We initiate the exploration of energetic advantages in driving operations of quantum systems obtained from non-thermal states. These energetic advantages with respect to initial thermal states are captured by the non-cyclic ergotropy, composed by the sum of a coherent (coherence-based) contribution, an incoherent (population-based) contribution and a passive contribution.
%which can be decomposed in three physically distinct contributions  %, which can be positive or negative. %More details???
%For a given initial state, the {\it non-cyclic ergotropy}, gives, when positive, the maximal amount of extractable energy, or conversely, when negative, the minimal energy investment required to complete the desired driving operation.
A more specific figure of merit can be introduced, $G_{\rho_i}$, focusing on the energetic gain brought by the optimal drive itself. It was shown to be equal to the non-cyclic ergotropy difference $\Delta {\cal E}_{\rm nc}$ for coherence-based non-passive states, as considered in the examples.
%, $G_{\rho_i}$, representing the energetic gains when applying optimal drives instead of adiabatic drives (or shortcut-to-adiabaticity) to initial non-passive states, was shown to be equal to the non-cyclic ergotropy difference $\Delta {\cal E}_{\rm nc}$ for coherence-based non-passive states.}

We saw that any non-thermal feature can bring energetic gains in the limit of many copies of the state. By contrast, for single state, we show, relying on majorization properties, that only quantum coherences can systematically bring energetic gains compared to initial thermal states. In particular, such gains are only achieved by dynamics able to consume coherences, emphasising the interplay between the presence of coherences and the ability to used them. It would be interesting to see if this mechanism could be the underlying phenomena behind the interferences effects enhancing the performance of cyclic engines pointed out in \cite{Camati_2019}, which would allow to extend its applications.
%We saw that non-thermal features based on coherences combined with coherent non-adiabatic drives -- able to consume coherences-- can always bring energetic gains. By contrast, non-thermal features based on populations do not always yield energetic benefits. %, and when they do, the energetic gains are accessible by incoherent non-adiabatic evolutions.

Additionally, the energetic costs associated with optimal drives can be significantly smaller than the ones associated with shortcut-to-adiabaticity technics while energetic gains can be significantly larger.
Future investigations should be conducted to analyse other systems and potentially other criteria to evaluate the energetic costs of drives \cite{Abah_2019}.

All these energetic advantages rely on the availability of non-thermal states. There are indeed many realistic situations producing non-thermal states, including strong interaction with a thermal bath \cite{Smith_2014,Nazir_2018}, many-body systems interacting with a common bath \cite{Benatti_2003,Muschik_2011,Krauter_2011,PRR2019b,Norcia_2018,minireview,PRA2020}, non-Markovian evolution \cite{Zhang_2013,Addis_2014}, and interaction with several thermal baths at different temperatures \cite{Leggio_2015}.
%a non-thermal bath. Notice that non-thermal baths can be effectively obtained form several thermal baths at different temperatures. Additionally, the solar radiation reaching us is indeed not in thermal state, providing a natural source of non-thermal bath. %, which are two situations of realistic experiments \cite{}. 
%(Additionally, even taking into account the energy necessary to generate the initial non-passive state out of a thermal state the energetic balance still might be positive). %SURE??????

Additional applications could be to explore questions suggested by our approach, like the lower energetic cost offered by the driving $V(t)$ compared to shortcut-to-adiabaticity, the tradeoff speed versus energetic cost of usual (cyclic) work extraction, as well as non-cyclic work extractions in quantum batteries. 
Finally, we anticipate direct applications in quantum engines operating with strong bath coupling \cite{Gelbwaser_2015,Newman_2017,Perarnau_2018,Newman_2020,Wiedmann_2020} or with structured bath \cite{Camati_2020}, where it has been reported mostly negative effects from the non-thermal properties and coherences naturally generated by these rich dynamics. A possible reason could be that such resources have not been fully exploited. Our results provide one possible direction.

\acknowledgements
This  work  is  based  upon  research  supported  by the National Institute for Theoretical Physics (NITheP) of the Republic of South Africa.

\appendix

\section{Incoherent and coherent non-adiabatic transformations}\label{appcoherent}
\subsection{Non-adiabatic transformations}
Any evolution leading to a passive state with respect to $H_f$ is necessarily non-adiabatic if and only if the initial state is a non-passive state with respect to $H_i$. This can be easily seen by writing the initial state in its diagonal form, $\rho_i=\sum_n r_n |r^i_n\ket \bra r^i_n|$. %where the unit vectors $|r_n\ket$ are the eigenvectors associated with the eigenvalues $r_n$ (decreasingly ordered).
 The final state is passive if and only if the applied evolution $U$ satisfies the condition $\bra e_n^f|U|r_{n'}\ket = \delta_{n,n'}$, where $\delta_{n,'n}$ is the Kronecker delta. Then, if $\rho_i$ is a passive state, we have $|r_{n'}\ket = |e_{n'}^i\ket$ for all $n'$, and the previous condition becomes $\bra e_n^f|U|e^i_{n'}\ket = \delta_{n,n'}$, which corresponds to an adiabatic transformation, or more precisely, to an ``integral or global" adiabatic transformation, a looser condition than a dynamics which is adiabatic at all intermediate times.  %(or to a shortcut-to-adiabaticity). 
 By contrast, if $\rho_i$ is not passive, it means there exists at least one $n$ such that $|r^i_n\ket \ne |e_n^i\ket$, implying that there exists at least another index $m\ne n$ satisfying $\bra e^i_{m}|r^i_n\ket \ne 0$. The condition for having a final passive state requires $\bra e_n^f|U|r_{n}\ket =1$, implying in fine that $U$ contains the transition $|e_n^f\ket\bra e_m^i|$, so that $U$ necessarily realises a non-adiabatic transformation.  
 %PUT THIS SHORT DEMO IN SM???

\subsection{Coherent and incoherent contributions}
 Non-thermal and non-passive features have two distinguished contributions: one from populations and one from coherences. It is possible to separate these two contributions in the non-cyclic ergotropy, extending similar considerations presented in \cite{Francica_2020} on ergotropy. However, for non-cyclic ergotropy, it is convenient to introduce a passive contribution, described in the following.
For an initial state $\rho_i$, we denote by $\rho_{i|_D}:= \sum_n \bra e_n^i|\rho_i|e_n^i\ket |e_n^i\ket\bra e_n^i|$ the corresponding dephased state. The incoherent contribution to the non-cyclic ergotropy is ${\cal E}_{\rm nc}^{\rm inc} := {\rm Tr} (\rho_i H_i ) - {\rm Tr}\left[ \widetilde{\rho_{i|_D}}   H_i\right]$, where $\widetilde{\rho_{i|_D}} = \sum_n p^i_{\sigma(n)} |e_n^i\ket\bra e_n^i|$ is the passive state of $\rho_{i|_D}$ with respect to $H_i$, with $p^i_n:=\bra e^i_n|\rho_i|e^i_n\ket$ are the populations in the initial energy basis, and $\sigma(n)$ is a permutation of the indices such that $p^i_{\sigma(n+1)} \leq p^i_{\sigma(n)}$. Obviously, in the particular situation where the populations $p_n^i$ of $\rho_i$ are already in decreasing order we have $\widetilde{\rho_{i|_D}}=\rho_{i|_D}$ and the incoherent contribution is null since $ {\rm Tr} (\rho_{i|_D} H_i )=  {\rm Tr} (\rho_i H_i )$. By defining the unitary transformation $U_{\sigma} := \sum_n e^{i\psi_n}|e^i_{n}\ket\bra e^i_{\sigma(n)}|$, where $\psi_n$ is a phase factor, we obtain $\widetilde{\rho_{i|_D}} = U_{\sigma}\rho_{i|_D} U_{\sigma}^{\dag} =  \left(U_{\sigma}\rho_{i} U_{\sigma}^{\dag}\right)_{|_D}$. Alternatively, $\widetilde{\rho_{i|_D}}$ can be defined as  \cite{Francica_2020} $\widetilde{\rho_{i|_D}} = {\rm argmin}_{\sigma \in {\cal S}^{\rm inc}} {\rm Tr}(\sigma H_i)$, where ${\cal S}^{\rm inc} := \{U_{\xi}\rho_{i|_D}U_{\xi}^{\dag}\}_{U_{\xi}\in{\cal U}^{\rm inc}}$ and ${\cal U}^{\rm inc}$ denotes the ensemble of incoherent unitary transformations with respect to the initial energy eigenbasis. Incoherent unitaries are of the form $\sum_n e^{i \Theta_n}|e^i_n\ket\bra e_{\xi(n)}^i|$, where $\xi(n)$ is a permutation of the indices and $\Theta_n$ a phase factor. Importantly, ${\cal E}_{\rm nc}^{\rm inc}$ is always positive. %Note that adiabatic transformations constitute the subgroup of incoherent unitary transformations with the permutation $\xi$ equal to identity. 

%The coherent contribution to the non-cyclic ergotropy can be defined as ${\cal E}_{\rm nc}^{\rm coh} = {\rm Tr} U_{\sigma}\rho_i U_{\sigma}^{\dag} H_f  - {\rm Tr}\widetilde{(\rho_i)}_f H_f $. The coherent contribution is always positive whereas the incoherent one can be of any sign. Both contributions add up to the non-cyclic ergotropy.  

The second contribution is the passive one, defined as ${\cal E}_{\rm nc}^{\rm pas} := {\rm Tr}\left[ \widetilde{ \rho_{i|_D}} H_i\right] - {\rm Tr}  \left[\widetilde{ (\rho_{i|_D})}_f H_f\right]$ with $ \widetilde{ (\rho_{i|_D})}_f:= \sum_n p^i_{\sigma(n)} |e_n^f\ket\bra e_n^f|$ is the passive state of $\rho_{i|_D}$ with respect to $H_f$. Note that $ \widetilde{ (\rho_{i|_D})}_f $ is related to $\widetilde{ \rho_{i|_D}}$ through adiabatic transformations which are of the form $U_{\rm ad}= \sum_n e^{i \phi_n} |e_n^f\ket\bra e_n^i|$.  Additionally, ${\cal E}_{\rm nc}^{\rm pas}$ can be positive or negative. %{\it What else?}

The coherent contribution to the non-cyclic ergotropy can be defined as ${\cal E}_{\rm nc}^{\rm coh} = {\rm Tr} \left[\widetilde{ (\rho_{i|_D})}_f H_f\right] - {\rm Tr}\left[\widetilde{(\rho_i)}_f H_f\right]$. ${\cal E}_{\rm nc}^{\rm coh}$ is always positive since ${\rm Tr}\left[ \widetilde{ (\rho_{i|_D})}_f H_f \right]= {\rm Tr} \left[U_{\rm ad}U_\sigma \rho_i U_\sigma^{\dag}U_{\rm ad}^{\dag} H_f\right]$ and $\widetilde{(\rho_i)}_f $ is the passive state associated with $\rho_i$ but also to $U_{\rm ad}U_\sigma \rho_i U_\sigma^{\dag}U_{\rm ad}^{\dag} $. A similar expression as in \cite{Francica_2020} can be obtained for the coherent non-cyclic ergotropy:
\bea
{\cal E}_{\rm nc}^{\rm coh} = \beta^{-1}\left[{\cal C}(\rho_i) + S[\widetilde{(\rho_{i|_D})}_f|\rho_f^{\rm th}(\beta)] - S[\widetilde{(\rho_i)}_f|\rho_f^{\rm th}(\beta)]\right]\nn\\
\eea
where $\rho_f^{\rm th}(\beta)$ denotes a thermal state of the final Hamiltonian at arbitrary inverse  temperature $\beta$, and ${\cal C}(\rho_i) = S[\rho_{i|_D}]-S[\rho_i]$ is the amount of initial coherences measured with the relative entropy of coherence \cite{Baumgratz_2014}. 
% It is different from zero only if the initial state contains coherences.
  While $\widetilde{ (\rho_{i|_D})}_f$ can be reached by incoherent non-adiabatic evolutions, for instance $U_{\rm ad}U_\sigma$, $\widetilde{ (\rho_{i})}_f$ can be reached only via coherent non-adiabatic evolutions. It can be easily seen by remembering that applying an incoherent non-adiabatic evolution to an initial state containing coherences necessarily yields a final state with coherences. 
Alternatively, one can see it by noticing that an evolution able to consume coherences is also able to generate coherences.

 %Here, $V_{\sigma}:= \sum_n e^{i\phi_n}|e^i_{n}\ket\bra e^i_{\sigma(n)}|$, which implies that only contribution to the non-passivity of $V_{\sigma}\rho_i V_{\sigma}^{\dag}$ is from coherences. 
%The coherent contribution is always positive whereas the incoherent one can be of any sign. 

Finally, the three contributions add up to give the non-cyclic ergotropy: ${\cal E}_{\rm nc} = {\cal E}^{\rm inc}_{\rm nc} + {\cal E}^{\rm pas}_{\rm nc} + {\cal E}^{\rm coh}_{\rm nc} $. Note that the passive contribution can alternatively be defined before the incoherent one or after the coherent one. This could in general change the respective value of each contribution but without changing their nature.  \\ %OBS: Note that other definitions are possible.   3 options for $??$, and in fact we also had three option for defining the incoherent contribution: with respect to $H_i$, with respect to $H_f$, or from $H_i$ to $H_f$.\\ 

\subsection{Consequences for energetic gain}\label{appcons}
%in terms of the different energetic contributions to the non-cyclic ergotropy introduced above and explicitly detailed in Appendix \ref{appcoherent}. 

The above decomposition of ${\cal E}_{\rm nc}$ is insightful to understand the difference between coherence and population-based non-passivity. 
For coherence-based non-passivity, the energetic gain, given by the difference of non-cyclic ergotropy (Eq.\ref{deltaE} of the main text), can be decomposed as 
\bea
\Delta {\cal E}_{\rm nc} &=&  {\cal E}_{\rm nc}(\rho_i)-{\cal E}_{\rm nc}(\rho_i^{\rm th})\nn\\
&=&{\cal E}_{\rm nc}^{\rm pas}(\rho_i) + {\cal E}_{\rm nc}^{\rm coh}(\rho_i) -  {\cal E}_{\rm nc}^{\rm pas}(\rho_i^{\rm th}).
%&=& {\cal E}_{\rm coh}(\rho_i) \geq 0,
\eea
%where ${\cal E}_{\rm pas}(\sigma)$ denotes the passive contribution to ${\cal E}_{\rm nc}$ associated to an initial state $\sigma$. 
Since we assumed that $\rho_i$ contains only coherence-based non-passive features, the populations are the same as the thermal state $\rho_i^{\rm th}$ of same energy, and consequently the passive contributions are also the same, ${\cal E}_{\rm nc}^{\rm pas}(\rho_i)={\cal E}_{\rm nc}^{\rm pas}(\rho_i^{\rm th})$. Consequently,
\bea
\Delta {\cal E}_{\rm nc} &=& {\cal E}_{\rm nc}^{\rm coh}(\rho_i) \geq 0,
\eea
 from which we conclude that coherences always bring energetic gains.
 
By contrast, for a population-based non-passive state, the populations can be very different from the thermal state of same energy, so that the passive contributions can be very different too. Then, we have 
\bea
\Delta {\cal E}_{\rm nc} &=&  {\cal E}_{\rm nc}^{\rm pas}(\rho_i) + {\cal E}_{\rm nc}^{\rm inc}(\rho_i) -  {\cal E}_{\rm nc}^{\rm pas}(\rho_i^{\rm th})\nn\\
&=&  \Delta{\cal E}_{\rm nc}^{\rm pas}+ {\cal E}_{\rm nc}^{\rm inc}(\rho_i),
\eea
which can be of any sign since $\Delta{\cal E}_{\rm nc}^{\rm pas}:= {\cal E}_{\rm nc}^{\rm pas}(\rho_i)  -  {\cal E}_{\rm nc}^{\rm pas}(\rho_i^{\rm th})$ can be positive or negative. This is illustrated in the next section with three-level systems.

\section{Larger non-cyclic work extraction from passive states than from non-passive states}\label{appexample}
In this Appendix we provide explicit example that for non-cyclic transformations, initial passive states can yield a larger work extraction (for positive non-cyclic ergotropy) or require less driving energy (for negative non-cyclic ergotropy) than non-passive states. We consider a three-level system and a non-cyclic process with initial Hamiltonian $H_i = \sum_{n=1}^3 e_n^i |e_n^i\ket\bra e_n^i|$ and the final Hamiltonian $H_f = \sum_{n=1}^3 e_n^f |e_n^f\ket\bra e_n^f|$. Without loss of generality, we assume that $e^i_1=e^f_1=0$ and $e^i_3=e^f_3 =1$, which implies that $e_2^i$ and $e_2^f$ belong to the interval $[0;1]$. As passive state, we consider a thermal state $\rho_i^{\rm th}$ at inverse temperature $\beta$. We denote by $p_n^{\rm th}$ its initial populations associated with the eigenvector $|e_n^i\ket$. We are looking for a non-passive state such that its non-cyclic ergotropy is strictly smaller than the one of the thermal state. Since we saw in the main text that initial coherences always increase the non-cyclic ergotropy, we choose a diagonal non-passive state $\rho_i  =\sum_{n=1}^3 q_i |e_n^i\ket\bra e_n^i|$. In other words, we need to find $q_1$, $q_2$, and $q_3$ such that $\widetilde E_f = q^{*}_3 + q^{*}_2 e_2^f >  {\rm Tr} \left[\widetilde{(\rho_i^{\rm th})}_f H_f\right]  = p_3^{\rm th} + p_2^{\rm th} e_2^f$, remembering that $\widetilde{(\rho_i^{\rm th})}_f $ is the final passive state associated with $\rho_i^{\rm th}$. We introduced $q_3^{*} := {\rm min}_{n=1,2,3}  ~ q_n$ and  $q_2^{*}$ is the second smallest population. 

One can show for instance that choosing $e_2^i$ such that $\beta(1-e_2^i)\ll1$, with $q_3 = p_3^{\rm th} + \alpha$, $q_2 = p_2^{\rm th} - \alpha/e_2^i$, and $q_1=1-q_2-q_3$, where $\alpha:=e^{-\beta}\beta(e_2^i-(e_2^i)^2)/3$ guarantees $q_3^{*}>p_3^{\rm th} $. This implies that we can always have $\widetilde E_f  >  {\rm Tr} \left[\widetilde{(\rho_i^{\rm th})}_f H_f\right]  $ by choosing $e_2^f$ small enough.

 Explicitly, let us take $\beta =1$ (in unit of $k_B$) and $e_2^i=0.9$. With these values we obtain according to the above choices $p_1^{\rm th} \simeq 0.564$, $p_2^{\rm th} \simeq 0.229$, $p_3^{\rm th} \simeq 0.207$, $q_1\simeq 0.565$, $q_2 \simeq 0.217$, and $q_3 \simeq 0.218$. Thus, we have for the final populations $q_3^{*} -p_3^{\rm th}= q_3 - p_3^{\rm th} \simeq 0.00954$ and  $p_2^{\rm th} - q_2^{*}  = p_2^{\rm th} - q_3 = 0.0108$, so that any value of $e_2^f$ smaller than $0.88$ leads to $\widetilde E_f  >  {\rm Tr} \left[\widetilde{(\rho_i^{\rm th})}_f H_f\right] $.

\section{Asymptotic achievability of the upper bound Eq.(7)}\label{appEq7}
 The theorem shown in \cite{Alicki_2013} states that for any state $\rho$ and for $N$ going to infinity, there exists a unitary transformation $U_N$ (not unique) such that 
\be
\frac{1}{N}{\rm Tr}\left( U_N \otimes^N \rho U_N^{\dag} H_N \right)\underset{N\rightarrow \infty}{\rightarrow} {\rm Tr} (\rho^{\rm th} H),
\ee
where $H$ can be an arbitrary Hamiltonian, $H_N:= \sum_{k=0}^{N-1}  \otimes^k\mathbb{I}\otimes H\otimes^{N-k-1}\mathbb{I} $, $\mathbb{I}$ is the identity, and $\rho^{\rm th}$ is the thermal state of same entropy as $\rho$ associated with the Hamiltonian $H$. Then, %independently of $H_i$, 
it implies the existence of a unitary transformation mapping asymptotically well (in the sense stated above) $\otimes^N\rho_i$ to $\otimes^N(\rho_i)_f^{\rm th}$, remembering that $(\rho_i)_f^{\rm th}$ denotes the thermal state with respect to $H_f$ of same entropy as $\rho_i$. In particular, it also means that we can find a time dependent Hamiltonian $H_N(t)$ such that the generated unitary transformation realises this mapping with the additional constraints $H_{N}(t_i) = \sum_{k=0}^{N-1}  \otimes^k\mathbb{I}\otimes H_i\otimes^{N-k-1}\mathbb{I} $ and $H_{N}(t_f)  = \sum_{k=0}^{N-1}  \otimes^k\mathbb{I}\otimes H_f\otimes^{N-k-1}\mathbb{I} $.

\section{Optimal drivings}\label{appdriving}
One can verify easily that the family of driving $V(t)$ given in the main text brings the initial state $\rho_i$ to the optimal final state $\tilde \rho_f$. The complete transformation is given by 
\bea
U &=&e^{-i{\cal T}\int_{t_i}^{t_f} dt H(t)}\nn\\
&=& e^{-i{\cal T}\int_{t_i}^{t_f} dt H_0(t)}\nn\\
&&\times e^{-i{\cal T}\int_{t_i}^{t_f} dt e^{i{\cal A}\int_{t_i}^tdu H_0(u)}V(t)e^{-i{\cal T}\int_{t_i}^tdu H_0(u)}}\nn\\
&=&e^{-i{\cal T}\int_{t_i}^{t_f} dt H_0(t)}e^{i\chi\int_{t_i}^{t_f} dt \dot f(t)}\nn\\
&=&e^{-i{\cal T}\int_{t_i}^{t_f} dt H_0(t)}e^{i\chi[f(t_f)-f(t_i)]}\nn\\
&=& \sum_n e^{i\phi_n}|e_n^f\ket\bra r^i_n|,
\eea 
where we used the definition of $\chi$ and the properties $f(t_i)=0$ and $f(t_f)=1$. It is then straightforward to see that $U$ brings $\rho_i$ to the optimal final state.

\section{Details on the energetic cost of driving for two-level systems}\label{appexTLS}
The eigenvalues and eigenvectors associated with the initial state $\rho_i = \begin{pmatrix}
    p_i & c_i \\
    c_i^{*} & 1-p_i \\
\end{pmatrix}$ are respectively 
\be
r_1=\frac{1}{2} - \sqrt{(p_i-1/2)^2 + |c_i|^2},
\ee
\be
r_0=\frac{1}{2} + \sqrt{(p_i-1/2)^2 + |c_i|^2},
\ee 
and
\be\label{r1}
|r_1\ket = \sqrt{\frac{r_0-p_i}{r_0-r_1}}|1\ket - e^{-i\psi_i}\sqrt{\frac{p_i-r_1}{r_0-r_1}}|0\ket,
\ee
\be\label{r0}
|r_0\ket = e^{i\psi_i}\sqrt{\frac{p_i-r_1}{r_0-r_1}}|1\ket + \sqrt{\frac{r_0-p_i}{r_0-r_1}}|0\ket,
\ee
where $\psi_i={\rm arg} (c_i)$. The unitary evolution $U_0(t)$ is simply given by $U_0(t) = e^{-i\Lambda\frac{\omega}{2}\sigma_z}$, with $\Lambda :=\int_{t_0}^{t_f}dt\lambda(t)$. Since both $H_i$ and $H_f$ are proportional to $\sigma_f$, the initial and final energy eigenstates are the same, namely $|1\ket$ and $|0\ket$.
As detailed in the main text, optimal drivings can be obtained through the matrix $\chi$ which is itself given by
\bea
e^{i\chi} &=& \sum_n e^{i\phi_n}U_0^{\dag}(t_f)|e^f_n\ket  \bra r^i_n|\nn\\
 &=&  e^{i\phi_1}e^{i\Lambda \frac{\omega}{2}}|1\ket  \bra r_1|+ e^{i\phi_0}e^{-i\Lambda \frac{\omega}{2}}|0\ket  \bra r_0|.
\eea   
The associated eigenvalues are $e^{i\theta_+}$ and $e^{i\theta_-}$ with 
\bea
\theta_{\pm}&=& \frac{\phi_0+\phi_1}{2}+\pi\kappa(\phi_1-\phi_0) \\
&&\pm {\rm arctan}\sqrt{\frac{r_0-r_1}{(r_0-p_i)\cos^2{(\phi_1-\phi_0)/2}}-1},
\eea
where $\kappa(\phi_1-\phi_0)$ is a function equal to 0 when $\cos(\phi_1-\phi_0)\geq 0$ and equal to 1 otherwise (explicitly, $\kappa(\phi_1-\phi_0) = \Theta[-\cos(\phi_1-\phi_0)]$, where $\Theta$ is the Heaviside step function).
Assuming one has full control of the phases, one can achieve the following minimal energetic cost 
\bea
w_{\rm min} &=&  \frac{\sqrt{2}}{\tau}{\rm arctan} \sqrt{\frac{p_i-r_1}{r_0-p_i}} \nn\\
&=&\frac{\sqrt{2}}{\tau}{\rm arctan}\left(\frac{\sqrt{(1/2-p_i)^2+|c|^2}-(1/2-p_i)}{\sqrt{(1/2-p_i)^2+|c|^2}+(1/2-p_i)}\right)^{1/2},\nn\\
\eea
by setting $\phi_1= \phi_0 = 0$.
By contrast, if one has no control of the phases, their value for each realisation is random, and the average cost is given by
\be
\bar w := \frac{1}{\tau}\frac{1}{4\pi^2}\int_{-\pi}^{\pi}d\phi_1\int_{-\pi}^{\pi}d\phi_2~ \sqrt{\theta_+^2+\theta_{-}^2}.
\ee  
%with $\overline{\theta_+^2}=\overline{\theta_-^2} = \overline{\theta^2}:= \frac{1}{4\pi^2}\int_{-\pi}^{\pi}d\phi_1\int_{-\pi}^{\pi}d\phi_2~ \theta^2_{\pm}$.
 The analytical expression is challenging to obtain, but one can instead show that $\frac{1}{4\pi^2}\int_{-\pi}^{\pi}d\phi_1\int_{-\pi}^{\pi}d\phi_2~ \sqrt{\theta_+^2+\theta_{-}^2}$ takes value within the interval $[0.77\pi; 0.89\pi]$ depending on the values of $p_i$ and $c_i$. \\

%{\it To take into account the phase $e^{\pm i\Lambda \frac{\omega}{2}}$, it is convenient to define new eigenvectors $|r_0\prime\ket $ and $|r_1\prime\ket$ as
%\be
%|r_1\prime\ket = e^{i\Lambda\frac{\omega}{2}}|r_1\ket =  \sqrt{\frac{r_0-p}{r_0-r_1}}|1\prime\ket - e^{-i\psi\prime}\sqrt{\frac{p-r_1}{r_0-r_1}}|0\prime\ket,
%\ee
%
%\be
%|r_0\prime\ket = e^{-i\Lambda\frac{\omega}{2}}|r_0\ket = e^{i\psi\prime}\sqrt{\frac{p-r_1}{r_0-r_1}}|1\prime\ket + \sqrt{\frac{r_0-p}{r_0-r_1}}|0\prime\ket,
%\ee
%with $\psi\prime = \psi -\Lambda \omega$, $|1\prime\ket = e^{i\Lambda\frac{\omega}{2}}|1\ket$ and $|0\prime\ket = e^{-i\Lambda\frac{\omega}{2}}|0\ket$. Then, it is easy to see that the eigenvalues of $e^{i\chi}$ are $e^{i \theta}$ and $e^{-i\theta}$, with 
%\be
%\theta = {\rm arctan} \frac{p-r_1}{r_0-p} ={\rm arctan}\left(\frac{\sqrt{1+\frac{|c|^2}{(1/2-p)^2}}-1}{\sqrt{1+\frac{|c|^2}{(1/2-p)^2}}+1}\right)^{1/2},
%\ee
%independent of the phase $\psi$ of the coherence, as expected since it is not a well-defined quantity.  
% Thus, the minimum energetic cost of the drive $V(t)$ is $w_{\rm min} = \sqrt{2}\theta/\tau$, achieved when $\dot f(t) \geq 0 \forall t \in [t_i;t_f]$, see main text.}

\section{Non-adiabatic dynamics}\label{appnonadiab}
The time dependent Hamiltonian considered in the last part of the paper is of the form
\be
H_0(t) = \frac{1}{2}\omega(t) \sigma_z + \frac{1}{2}\epsilon(t)\sigma_x,
\ee
and generates a non-adiabatic dynamics since $[H(t),H(t')]\ne 0$ in general. Such kind of dynamics are challenging to integrate. Still, analytical integrations are possible when the Hamiltonian parameters are such that $\mu := \frac{\dot \omega(t) \epsilon(t) - \dot \epsilon(t) \omega(t)}{\Omega^3(t)}$ is constant \cite{Feldmann_2003, Dann_2020}. A simple way to integrate the dynamics is using a closed set of orthogonal observables $\{B_k\}_{0\leq k\leq 3}$ forming a basis of the Hilbert space. We use the same set as in \cite{Feldmann_2003, Dann_2020}, namely $B_0= \mathbb{I}$, $B_1=H_0(t) = \omega(t) S_z+\epsilon(t) S_x$, $B_1=\epsilon(t) S_z - \omega(t) S_x$, and $B_3=\Omega(t) S_y$. In the Heisenberg picture $B_i(t):=e^{i{\cal A}\int_{t_i}^{t} duH_0(u)} B_i  e^{-i{\cal T}\int_{t_i}^{t} duH_0(u)}$ the dynamics is given by $\dot B_i(t) = i[H_0^{*}(t),B_i(t)] + \frac{\partial}{\partial t}B_i(t)$, where $H_0^{*}(t):=e^{i{\cal A}\int_{t_i}^{t} duH_0(u)} H_0  e^{-i{\cal T}\int_{t_i}^{t} duH_0(u)}$ and the partial derivative denotes the derivative with respect to the intrinsic time-dependence of the operator $B_i$. The dynamics of the basis can be written in a matrix form
\be
\frac{1}{\Omega} \dot X(t) = \left(A+\frac{\dot\Omega}{\Omega^2}\mathbb{I}\right)X(t)
\ee 
where $X(t)=\{B_1(t),B_2(t),B_3(t)\}^{\rm T}$ is a three-component column vector and 
\be
A = \begin{pmatrix}
    0 & \mu & 0 \\
   -\mu & 0 &1\\
   0 & -1 & 0
\end{pmatrix}
\ee 
a $3\times3$-matrix. This can be integrated after diagonalising $A$, yielding (see also \cite{Feldmann_2003, Dann_2020})
\begin{widetext}
\be
X(t) = \frac{\Omega(t)}{(\mu^2+1)\Omega(0)}\begin{pmatrix}
    1+\mu^2 \nu_c & \mu \nu_s\sqrt{\mu^2+1} & (1-\nu_c)\mu \\
   -\mu \nu_s\sqrt{\mu^2+1}& \nu_c(\mu^2+1) &\nu_s\sqrt{\mu^2+1}\\
   \mu(1-\nu_c) & -\nu_s\sqrt{\mu^2+1} & \mu^2+\nu_c
\end{pmatrix} X(0),
\ee
\end{widetext}
with $\nu_c:=\cos \bar\Omega\sqrt{1+\mu^2}$, $\nu_s:=\sin \bar\Omega\sqrt{1+\mu^2}$, and $\bar\Omega := \int_{t_i}^{t_f} dt \Omega(t)$. Note that from $B_i(t)$ we have directly the expressions of $S_z(t)$, $S_y(t)$, and $S_x(t)$, from which we obtain the time dependent Bloch vector to reconstruct the final state, $\rho_f = \begin{pmatrix}
   p_f& c_f \\
    c_f^{*}& 1-p_f\\
\end{pmatrix}$ in the basis $\{|e_1^f\ket,|e_0^f\ket\}$. We obtain $p_f= \frac{1}{2(1+\mu^2)}[2p_i + \mu^2 -\mu^2\nu_c(1-2p_i)]$ and $c_f=- \frac{\mu(1-2p_i)}{2(1+\mu^2)}{\rm sign}(\epsilon_f)[\nu_s\sqrt{1+\mu^2} - i(1-\nu_c)]$. The final eigenstates are given by $|e_1^f\ket = [(\omega_f+\Omega_f)^2+\epsilon_f^2]^{-1/2}[(\omega_f+\Omega_f)|1\ket +\epsilon_f |0\ket]$ and $|e_0^f\ket = [(\omega_f-\Omega_f)^2+\epsilon_f^2]^{-1/2}[(\omega_f-\Omega_f)|1\ket +\epsilon_f |0\ket]$.

\subsection{Energetic cost of the family of optimal drive}\label{appfamily}
The minimal energetic cost of an optimal drive is given by (see main text) $w_{\rm min} = \frac{1}{\tau} [{\rm Tr} \chi \chi^{\dag}]^{1/2}$, where $\chi$ is such that $e^{i\chi} = \sum_ne^{i\xi_n}|e_n^i\prime\ket\bra r^i_n|$. The expression of $|r_1\ket$ and $|r_0\ket$ are the same as in \eqref{r1} and \eqref{r0} respectively. We can derive the expression of $|e_1^i\prime\ket$ and $|e_0^i\prime\ket$ (up to a phase factor included in $\xi_n$) from the above expression of $\rho_f$, taking respectively $ |1\ket\bra 1|$ and $|0\ket\bra 0|$ as initial state. We obtain
\be\label{e1}
|e_1^i\prime\ket = \alpha e^{-i\phi_{\alpha}} |1\ket - \beta|0\ket
\ee
\be\label{e0}
|e_0^i\prime\ket = \beta |1\ket + \alpha e^{i\phi_{\alpha}}\beta|0\ket
\ee
with $\alpha e^{i\phi_{\alpha}} = \frac{1}{\sqrt{2(1+\mu^2)}}[{\rm sign}(s)\sqrt{(1+\mu^2)(1+\nu_c)}-i\sqrt{1-\nu_c}]$ and $\beta =\frac{1}{\sqrt{2(1+\mu^2)}}{\rm sign}(\epsilon_f)\mu\sqrt{1-\nu_c}$.

Combining with \eqref{e1} and \eqref{e0} with \eqref{r1} and \eqref{r0} we have
\bea
\sum_n e^{i\phi_n} |e_n^i\prime\ket \bra r^i_n| =  \begin{pmatrix}
   \cos \eta e^{i\zeta} & \sin\eta e^{-i(\delta-\xi_1-\xi_0)} \\
    -\sin\eta e^{i\delta}& \cos\eta e^{-i(\zeta-\xi_0-\xi_1)}\\
\end{pmatrix}\nn\\
\eea
in the initial energy eigenbasis $\{|1\ket,|0\ket\}$ with $\eta$, $\zeta$, and $\delta$ are implicitly defined by the relations $\cos \eta e^{i\zeta} =\alpha \sqrt{\frac{r_0-p_i}{r_0-r_1}}e^{i(\xi_1-\phi_\alpha)} + \beta\sqrt{\frac{p_i-r_1}{r_0-r_1}}e^{i(\xi_0-\psi_i)}$ and $\sin\eta e^{i\delta} = \beta \sqrt{\frac{r_0-p_i}{r_0-r_1}}e^{i\xi_1} - \alpha \sqrt{\frac{p_i-r_1}{r_0-r_1}}e^{i(\xi_0+\phi_\alpha -\psi_i)}$,
reminding that $\psi_i$ is the argument of the initial coherence $c_i$.
As a result, the eigenvalues of $e^{i\chi}$ are $e^{i\theta_+}$ and $e^{i\theta_-}$ with
\begin{widetext}
\bea
\theta_{\pm} &=& \frac{\xi_1+\xi_0}{2} + \pi\kappa(\xi_1,\xi_0,\alpha,\eta) \pm {\rm arctan}\sqrt{\frac{1}{\left[\alpha\sqrt{\frac{r_0-p_i}{r_0-r_1}}\cos(\phi_\alpha-\xi_1/2+\xi_0/2)+\beta\sqrt{\frac{p_i-r_1}{r_0-r_1}}\cos(\psi_i+\xi_1/2-\xi_0/2)\right]^2}-1},\nn\\
\eea
where $\kappa(\xi_1,\xi_0,\alpha,\eta) $ is a function equal to 0 when $\cos\eta\cos(\zeta-\xi_1/2-\xi_0/2) \geq0$, and equal to 1 otherwise. Assuming one has control of the phase $\xi_1$ and $\xi_0$, the minimum energetic cost is 
\be
w_{\rm min} = \frac{\sqrt{2}}{\tau}{\rm arctan}\sqrt{\left[\alpha^2\frac{r_0-p_i}{r_0-r_1}+\beta^2\frac{p_i-r_1}{r_0-r_1} +2\beta\alpha\frac{\sqrt{(r_0-p_i)(p_i-r_1)}}{r_0-r_1}\cos(\psi_i+\phi_\alpha)\right]^{-1}-1}.
\ee
\end{widetext}
Contrasting with the first situation where $H_0(t)$ generates an adiabatic transformation, the energetic cost after minimisation over $\xi_1$ and $\xi_0$ depends on $\psi_i$, the initial phase of the coherence $c_i$, and on $\phi_{\alpha}$ (which depends on the original dynamics $U_0(t_f)$). Then, one can consider again the same alternative. If experimentally one has control of these phases, meaning that they have well-defined values which can be adjusted by some controls on the experimental apparatus, then, the minimum energetic cost can be brought down to (for $\psi_i+\phi_\alpha=0$ if $\epsilon_f\mu\geq0$, and for $\psi_i+\phi_\alpha=\pi$ if $\epsilon_f\mu\leq0$)
\bea
w_{\rm min} %&=& \frac{\sqrt{2}}{\tau}{\rm arctan}|\tan(\theta_{1,{\rm min}} -\theta_{2,{\rm min}})| \nn\\
&=&\frac{\sqrt{2}}{\tau}|\theta_{1,{\rm min}} -\theta_{2,{\rm min}}|,
\eea
with $\theta_{1,{\rm min}} = {\rm arctan} \sqrt{\frac{p_i-r_1}{r_0-p_i}} $, contribution form the initial state, and $\theta_{2,{\rm min}} = {\rm arctan} \frac{|\beta|}{\alpha} = {\rm arctan}\frac{|\mu|\sqrt{1-\nu_c}}{\sqrt{2+\mu^2(1+\nu_c)}}$, contribution from the original dynamics $U_0(t_f)$. Conversely, if one has no control over $\psi_i$ and $\phi_{\alpha}$, the energetic cost can take any value between $\frac{\sqrt{2}}{\tau}|\theta_{1,{\rm min}} -\theta_{2,{\rm min}}|$ and $\frac{\sqrt{2}}{\tau}(\theta_{1,{\rm min}} +\theta_{2,{\rm min}})$.

Finally, without any control one the phases $\xi_1$ and $\xi_0$ and therefore left random, the energetic cost takes values between $\frac{\sqrt{2}}{\tau}|\theta_{1,{\rm min}} -\theta_{2,{\rm min}}|$ and $\frac{\sqrt{2}}{\tau}\pi$.

\subsection{Counterdiabatic driving and energetic cost}\label{appcostSTA}
According to \cite{Demirplak_2003, Demirplak_2005, Demirplak_2008, Berry_2009}, for a given time dependent Hamiltonian $H(t)$, the counterdiabatic drive is given by
\be
H_{\rm CD} = \sum_n \left(\frac{d}{dt}\pi_n\right)\pi_n
\ee
where $\pi_n :=|e_n\ket\bra e_n|$ are projectors onto the instantaneous eigenstates $|e_n\ket$ of $H(t)$. The corresponding energetic cost is given by the time average of the Hamiltonian norm $||H(t)||$. One obtains $||H(t)|| = \big(\sum_n \dot{\bra e_n}\dot{|e_n\ket}\big)^{1/2}$, using the property $\dot{\bra e_n|}|e_n\ket=0$, where $\dot{|e_n\ket}$ stands for $\frac{d}{dt}|e_n\ket$. Recalling that we consider the family of driving $H_0(t) = \frac{\omega(t)}{2}\sigma_z+\frac{\epsilon(t)}{2} \sigma_x$, the instantaneous eigenstates are given by
\be
|e_1\ket = \sqrt{\frac{\Omega(t) + \omega(t)}{2\Omega(t)}} |1\ket + \sqrt{\frac{\Omega(t) - \omega(t)}{2\Omega(t)}} |0\ket
\ee
and
 \be
|e_0\ket = -\sqrt{\frac{\Omega(t) - \omega(t)}{2\Omega(t)}} |1\ket + \sqrt{\frac{\Omega(t) +\omega(t)}{2\Omega(t)}} |0\ket.
\ee
This leads to $||H(t)|| = \frac{|\dot\omega(t)\epsilon(t)-\omega(t)\dot\epsilon(t)|}{\Omega^2(t)} = |\mu| \Omega(t)$, implying that the energetic cost is $w_{\rm STA} = |\mu|\bar\Omega/\tau$.

\begin{figure}
\centering
(a)\!\!\!\!\!\!\!\!\!\!\!\!\!\!\!\!\includegraphics[width=8.5cm, height=4cm]{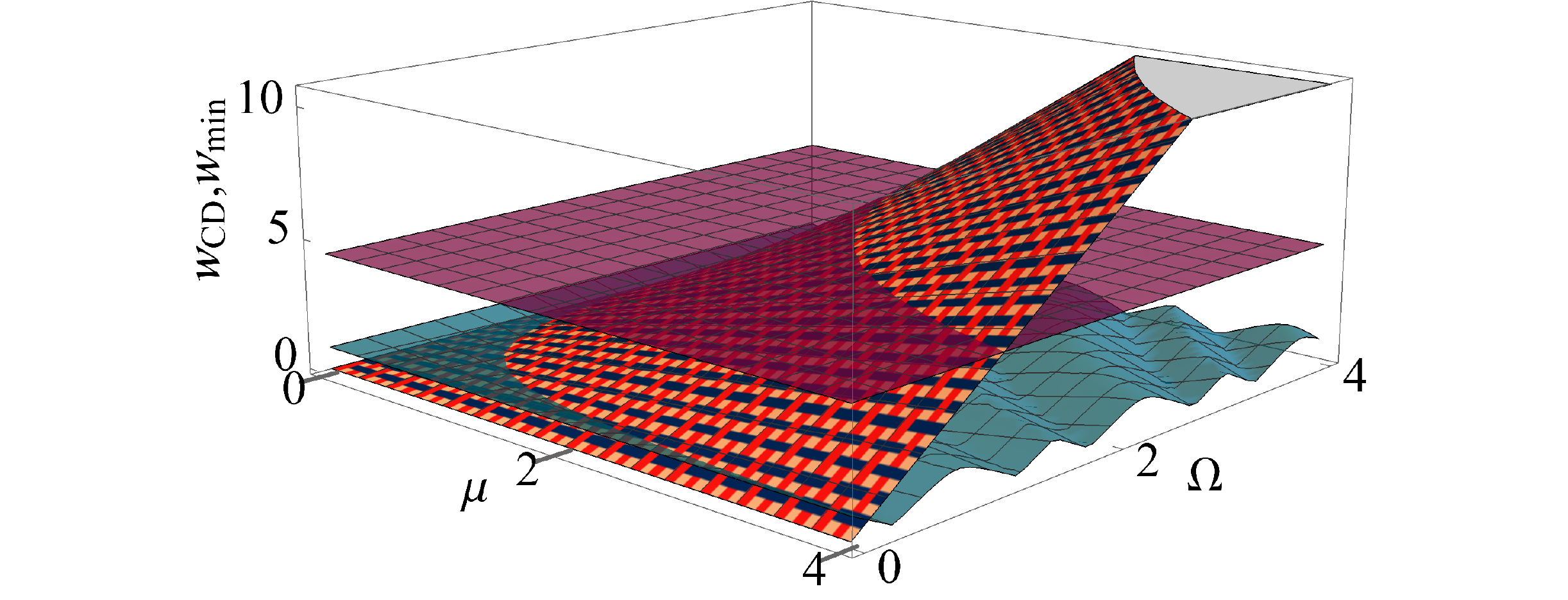}\\
(b)\!\!\!\!\!\!\!\!\!\!\!\!\!\!\!\!\!\!\!\!\!\!\!\!\!\!\!\!\includegraphics[width=6.5cm, height=4.4cm]{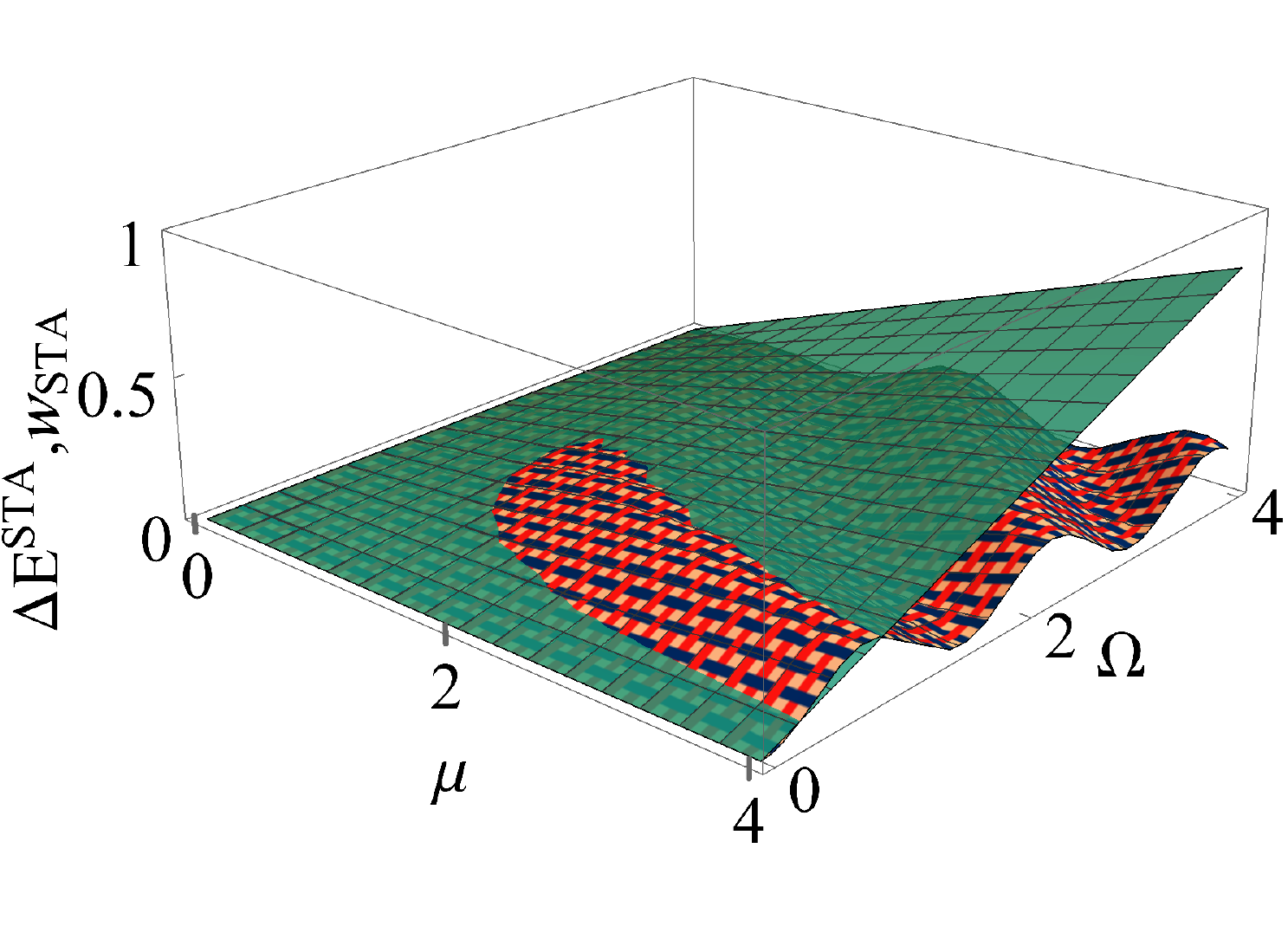}\\
(c)\!\!\!\!\!\!\!\!\!\!\!\!\!\!\!\!\!\!\!\!\!\!\!\!\!\!\!\!\includegraphics[width=7.5cm, height=4cm]{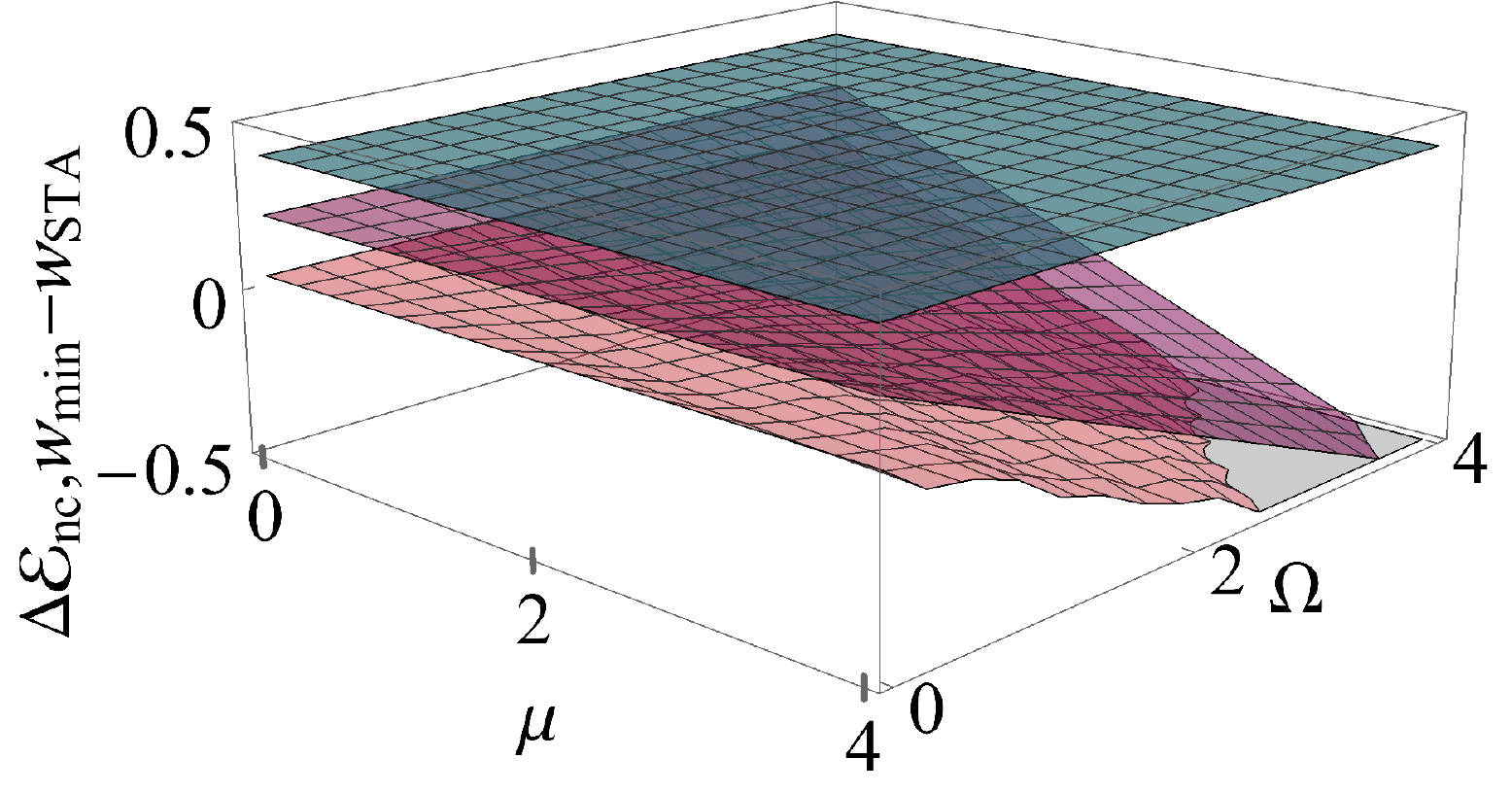}
\caption{(a) Plot of $w_{\rm STA}$ (texturised surface) and $w_{\rm min}$ (blue solid lower surface) in unit of $\hbar/\tau$, in function of the dimensionless parameters $\mu \in [0;4]$ and $\bar\Omega \in [0;4]$, for $p_i=0.4$ and $c_i = \sqrt{0.4\times0.6}$. The purple upper plane is the upper bound of $w_{\rm min}$ equal to $\pi\sqrt{2}$. (b) Plot of $w_{\rm STA}$ (green solid surface) and $\Delta E^{\rm STA}$ (texturised surface), both in unit of $\hbar\Omega_f$, in function of $\mu \in [0;4]$ and $\bar\Omega \in [0;4]$, assuming $\Omega_f = 20/\tau$, which allows to convert $\hbar/\tau$ in unit of $\hbar\Omega_f$. (c) Plot of $w_{\rm min} -w_{\rm STA}$ (pink lower surface) and $\Delta {\cal E}_{\rm nc}$ (upper blue horizontal plane), both in unit of $\hbar\Omega_f$, in function of $\mu \in [0;4]$ and $\bar\Omega \in [0;4]$, still assuming $\Omega_f = 20/\tau$. The purple intermediate surface is the upper bound of $w_{\rm min}-w_{\rm STA}$, equal to $\pi\sqrt{2}-w_{\rm STA}$.}
\label{comparison}
\end{figure}

\subsection{Some additional plots}\label{appaddplots}
We finally provides some plots additional to the one presented in the main text. The following plots are in function of the more general parameters $\overline{\Omega}$ and $|\mu|$. 
In Fig. \ref{comparison} (a), we compare the energetic cost $w_{\rm min}$ and $w_{\rm STA}$ in unit of $\hbar/\tau$ and in function of the dimensionless parameters $\mu \in [0;4]$ and $\bar\Omega \in [0;4]$, and setting $p_i=0.4$ and $c_i = \sqrt{0.4\times0.6}$.
One can see that $w_{\rm min}$ is almost always smaller than $w_{\rm STA}$. Additionally, for some initial non-passive states, the energetic cost $w_{\rm min}$ is zero, meaning that the transformation $U_0(t_f)$ is already optimal for these particular initial states. 
 Without the phase controls mentioned above, $w_{\rm min}$ takes random values between $\frac{\sqrt{2}\hbar}{\tau}|\theta_{1,{\rm min}} - \theta_{2,{\rm min}}|$ and $\frac{\sqrt{2}\hbar}{\tau}\pi$, so one could say that on average the energetic costs $w_{\rm min}$ and $w_{\rm STA}$ are comparable for moderate values of $\bar \Omega$ and $\mu$. For large values of these parameters, the $w_{\rm min}$ is always smaller than $w_{\rm STA}$. 

In Fig. \ref{comparison} (b) we compare the energetic gain $\Delta E^{\rm STA}$ in unit of $\hbar \Omega_f$ brought by shortcut-to-adiabaticity with its energetic cost $w_{\rm STA}$ in unit of $\hbar/\tau$. Since these two parameters are in principle independent, in order to be able to plot these two functions on the same graph we have to fixed a ``conversion rate" of $\hbar\Omega_f$ into $\hbar/\tau$. We choose $\Omega_f = 20/\tau$.
 One can see that the energetic balance is negative for most parameter values. In Fig. \ref{comparison} (c), using the same unit, we compare the energetic gain $\Delta {\cal E}_{\rm nc}={\cal E}_{\rm nc} - {\cal E}_{\rm nc}^{\rm STA} $ brought by the initial coherences with the relative energetic cost $w_{\rm min}-w_{\rm STA}$, still assuming $\Omega_f = 20/\tau$. 
 Overall, it seems that the performances of the optimal drive are significantly better than the performances of shortcut-to-adiabaticity.

\end{document}